\documentstyle[10pt,aaspp4,flushrt]{article}

\begin{document}

%

~~~

\vskip1truein

\title{MORPHOLOGICAL EVOLUTION OF GALAXIES}

\author{Hugo Martel,\altaffilmark{1} Premana Premadi,\altaffilmark{2,3,4} 
and Richard Matzner\altaffilmark{2,3,5}}

\altaffiltext{1}{Department of Astronomy, University of Texas, Austin, TX 
78712}
\altaffiltext{2}{Center for Relativity, University of Texas, Austin, TX 78712}
\altaffiltext{3}{Department of Physics, University of Texas, Austin, TX 78712}
\altaffiltext{4}{Current Address: Astronomical Institute, Tohoku University,
Sendai, Japan}
\altaffiltext{5}{Orson Anderson Scholar, Los Alamos National Laboratory 
1996-97}

\begin{abstract}
We simulate the growth of large-scale structure, for 3 different 
cosmological models, an Einstein-de Sitter model (density parameter
$\Omega_0=1$), an open model ($\Omega_0=0.2$) and a flat model 
with nonzero cosmological constant
($\Omega_0=0.2$, cosmological constant $\lambda_0=0.8$),
using a cosmological N-body code ($\rm P^3M$) with $64^3$ dark matter particles
in a comoving cubic volume of present comoving size 128 Mpc. 
The calculations start at $z=24$ and end at $z=0$.
We use the results of these simulations to generate distributions of
galaxies at the present ($z=0$), as follows: Using a Monte-Carlo method
based on the present distribution of dark matter,
we located $\sim40000$ galaxies in the computational volume.
We then ascribe to each galaxy a morphological type based on
the local number density of galaxies in order to reproduce
the observed morphology-density relation. The resulting galaxy distributions 
are similar to the observed ones, with most ellipticals concentrated in 
the densest regions, and most spirals concentrated in low-density regions.
By ``tying'' each galaxy to its nearest dark matter particle,
we can trace the trajectory of that galaxy back in time, by simply
looking at the location of that dark matter particle at earlier
time-slices provided by the N-body code. This enables us to
reconstruct the distribution of galaxies at high redshift, and the trajectory
of each galaxy from its formation epoch to the present.

We use these 
galaxy distributions to investigate the problem of morphological evolution.
Our goal is to determine whether the morphological type of galaxies
is primarily determined by the initial conditions in which
these galaxies form, or by evolutionary processes 
(such as mergers or tidal stripping) occurring after the galaxies have formed,
and eventually altering their morphology, or a combination of both effects. 
Our main technique consists of
comparing the environments in which galaxies are at the epoch
of galaxy formation (taken to be at redshift
$z=3$) with the environment in which
the same galaxies are at the present. Making the null hypothesis that 
the morphological types of galaxies
do not evolve, we compare the galaxies that form in low density 
environments but end up later in high density environment to the ones
that form also in low density environment but remain in low density 
environment. The first group contains a larger proportion of elliptical and
S0 galaxies than the second group. 
We assume that the initial galaxy formation process 
cannot distinguish a low density environment that will always
remain low density from one that will eventually become high density.
Therefore, these
results are absurd and force us to discard the null hypothesis that
morphological evolution does not occur. Our study suggests that $\sim75\%$
of the elliptical and S0 galaxies observed at present formed as such,
while the remaining $\sim25\%$ of these galaxies formed as spiral
galaxies, and underwent morphological evolution, for all three cosmological
models considered (the percentages might be smaller for elliptical than
S0 galaxies). These numbers assume a morphological evolution 
process which converts one spiral galaxy into either a S0 or an elliptical
galaxy. If the morphological evolution process
involves mergers of spiral galaxies, 
these numbers be would closer to $85\%$ and $15\%$, respectively.
We conclude that most galaxies did not undergo morphological evolution,
but a non-negligible fraction did.
\end{abstract}


%

\section{INTRODUCTION}

\subsection{Morphological Types}

Galaxies exist in several forms, elliptical, lenticulars, spirals,
and irregulars, usually referred to as {\it morphological types}.
Elliptical galaxies are featureless, ellipsoidal
stellar systems composed of old
Population II stars, with no appreciable amount of cold gas or dust.
In addition, many of them are known to contain also a disk.
Ellipticals galaxies are labeled as E0, E1, and so on,
according to their ellipticity.
Lenticular galaxies have a prominent, featureless disk, that contains no 
appreciable amount of cold gas or dust, and no spiral arms. 
They are very similar to the most elongated, E7 elliptical galaxies.
These galaxies are labeled as S0. Spiral galaxies are composed of a disk of
Population~I stars, cold gas, and dust, arranged in a pattern
of spiral arms,
and a central bulge of population~II stars which resemble small 
elliptical galaxies. The spiral arms are the site of active star formation, 
and contain a large number of young O and B stars.
Spiral galaxies have flat rotation curves that extend to radii well beyond the 
visible edge of the galaxy, thus implying that these galaxies
are imbedded into large dark matter halo. Spiral galaxies are
labeled as Sa, Sb, Sc, and Sd galaxies according to their disk-to-bulge
luminosity ratio (D/B), 
with the bulge dominating the luminosity for Sa galaxies,
and the disk dominating for Sd galaxies. Galaxies that do not belong to any
of these categories are classified as irregular galaxies. Some irregular
galaxies result from collision and merging between galaxies, but the majority
of irregular galaxies are small, gas rich galaxies similar to the
Magellanic clouds. We label these galaxies as Im.

All the galaxy types described above can be combined into a single sequence, 
$\rm E0\rightarrow E1\rightarrow\ldots\rightarrow E7
\rightarrow S0\rightarrow Sa\rightarrow Sb
\rightarrow Sc\rightarrow Sd\rightarrow Im$, called the {\it Hubble sequence}
\footnote{The Hubble sequence is actually a ``tuning fork'' with
two branches, one for {\it unbarred} spirals and one for
{\it barred} spirals. In this paper, we ignore the difference between barred
and unbarred spirals, thus collapsing the tuning fork into a rod. Hence,
``Sa'' designates both Sa and SBa galaxies, and so on.}.
Near the start of the sequence, galaxies are mostly composed of old
Population~II stars, with no dust and no cold gas, and therefore no active
star formation, and a small disk-to-bulge ratio. 
As we move along the sequence, the preponderance of Population~II stars
decreases in favor of younger, Population~I stars. The amount
of dust and cold gas increases, D/B
increases, and star formation becomes important.

A successful theory of galaxy formation must be able to explain the 
existence of the Hubble sequence, the origin of each morphological
type, their relative abundance,
and their clustering properties. To achieve this goal, we must
first identify and understand the physical processes that are involved in 
the galaxy formation process, as well as the processes that might 
subsequently alter the structure of galaxies after they are formed.
The most important clue for understanding the galaxy formation process is
the existence of a {\it Morphology-Density Relationship} relating the 
likelihood of any given galaxy to have a particular morphological type
to the {\it local} density of the environment in which that galaxy is located.

\subsection{The Morphology-Density Relation at Present}

There is a significant difference between the galaxy populations
of nearby low-density fields and in the densest regions inside nearby
clusters of galaxies. Though all morphological types are present
both in clusters and in the field, field galaxies are predominantly
spirals, while clusters of galaxies contain a much larger proportion
of elliptical and S0 galaxies. Furthermore, population gradients are found
inside clusters. Melnick \&~Sargent (1977) showed that the proportion of
spiral galaxies increases as a function of the distance from the cluster 
center, with a corresponding decrease in the proportion of S0 and 
elliptical galaxies. Dressler (1980) argued that this morphology-radius
relation is applicable only to regular, spherical clusters
with a well-defined center. Most clusters are highly irregular, and often
contain several high density concentrations, or lumps. The distribution of the
various morphological types inside these lumps is similar to the one in
the center of the regular, spherical clusters. Dressler (1980) concluded
that the correct way to describe the distribution of morphological types
is in terms of the local number density of galaxies, and not the distance
from the cluster center. Using a sample of 55 rich clusters, he showed that
the fraction of elliptical and S0 galaxies increases with increasing
surface number density of galaxies, with a corresponding decrease in
the fraction of spiral galaxies, over 3 orders of magnitude in surface
number density. The lowest density regions in the sample
are composed of 80\% spirals, 10\% S0's, and 10\% 
ellipticals, while the densest clumps are composed of 10\%
spirals, 40\% S0's, and 50\% ellipticals. Subsequent studies 
(Bhavsar 1981; de~Souza et al 1982; Postman \& Geller 1984)
confirmed the relations derived by Dressler (1980), and extended 
them to the low-density field. All these results are summarized in 
Dressler (1984). The morphology-density relation extends over
5 orders of magnitude in volume number density (Postman \&~Geller claim
6 orders of magnitude), and is a slowly varying, monotonic relation.
The lowest-density regions are composed of 80--90\% spirals, while
the highest-density regions are composed of 80--90\% ellipticals and
S0's. (Notice that a recent paper by Whitmore, Gilmore, \&~Jones
[1993] challenges the existence of the morphology-density relation,
and claims that the morphology-radius relation is actually the correct one.)

Notice that these various determinations of the morphology-density relation
were all based on observations of relatively nearby galaxies. Therefore,
this relation is valid only {\it at present}. More recent observations
with the Hubble Space Telescope (HST) suggest that the morphology-density 
relation evolves with time, and this actually supports the results we present 
in this paper. A discussion of the HST results is presented in \S9.

\subsection{The Origin of the Morphological Types}

Several galaxy formation models have been suggested to explain the origin
of the Hubble sequence and the existence of the morphology-density
relation. Dressler (1984) has grouped these various models into three
classes, based on the relative importance of initial conditions and evolution 
processes in determining the final morphological type of galaxies.
We shall follow the same classification here.

\subsubsection{Morphological Evolution}

Models that belong to the first class all assume that galaxies form
in similar environments, and therefore the existence of different
morphological types does not result from different initial conditions,
but instead from evolutionary processes happening after the galaxies have 
formed. Several models have been suggested to explain the abundance of S0
galaxies and deficiency of spiral galaxies in dense regions.
These models all assume that S0 galaxies are spiral galaxies that have lost 
their gas and dust as a result of some evolutionary process taking place
in the dense environments of cluster cores. The various possible
physical mechanism for gas stripping include direct collisions
(Spitzer \&~Baade 1951) ram-pressure stripping (Gunn \&~Gott 1972) and gas
evaporation by a hot intracluster gas (Cowie \& Songaila 1977). Dressler
(1980) pointed out a major problem with these models: the various physical
mechanisms suggested are efficient only in the densest regions, inside
cluster cores. Though the {\it fraction} of S0 galaxies is largest in these
regions, the {\it actual number} of S0's galaxies in these regions is
small. About 80\% of S0 galaxies are located in intermediate-density regions.
Spiral galaxies in intermediate-density 
regions are deficient in gas by a factor of 2-3
relative to field spirals, indicating that gas ablation is important in
these regions as well (Giovanelli, Chincarini, \& Haynes 1981;
Bothun, Schommer, \&~Sullivan 1982; 
Kennicutt 1983). However, this effect is much too weak 
to explain the presence of S0 galaxies, which are gas deficient by a factor
of 100 relative to field spirals.

\subsubsection{Initial Conditions Combined with Morphological Evolution.}

The second class of models comprises all models in which both 
initial conditions and morphological evolution play an important
role in determining the morphological types of galaxies. Kent (1981) had
suggested that the morphology-density relation originates from the ``fading''
of disks in high density regions. In this model, initial conditions are
assumed to be responsible for determining the initial morphological type
of disk galaxies, such that disk galaxies with large D/B become predominantly
spirals, while disk galaxies with small D/B become predominantly
S0's. The model then assumes that the disks of spiral and S0 galaxies
are fainter in high density regions than in low density region (this could
result from the dissipation of the disk by tidal interaction, or, if the
disks are still in the process of forming, then a large density environment
might disrupt this process). The fading of disks causes some
spiral galaxies to become too faint to be observable, and others to be
identified as S0 galaxies. Furthermore, the fading of the disk of S0
galaxies causes some of these galaxies to be identified as ellipticals.
With an appropriate choice of parameters, this model can successfully
reproduce all the relations given in Dressler (1980). 
Larson, Tinsley, \&~Caldwell (1980) have proposed a similar model, based on
the time scale for gas exhaustion via stellar evolution in disks. In their
model, the gas exhausted by star formation is constantly replaced by gas
infalling from a gaseous envelope surrounding the galaxy. In high-density
regions, tidal encounters would disrupt this envelope, resulting in a 
progressive fading of the disk as stellar evolution proceeds.
The various gas-stripping processes mentioned in \S1.3.1 could be responsible 
for transforming spirals galaxies into S0's inside cluster cores (even
though they cannot account for the existence of field S0 galaxies). 
Byrd \&~Valtonen (1990) have argued that the interaction of spiral galaxies
with the tidal field of the cluster is a more efficient process
than ram pressure stripping in depleting these galaxies of their 
interstellar gas, and eventually turning them into S0 galaxies
(but not ellipticals). Their model is supported by the abundance
of barred spiral galaxies in the core of the Coma cluster, since the
formation of a bar in a normal spiral galaxy can also result from 
strong tidal interaction.

If the galactic disks are ``faded'' in high density regions, as these 
models assume,
then the luminosity function inside dense clusters should differ
significantly from the one in low density clusters and in the field.
However, observations show that the luminosity functions in low-
and high-density regions are essentially identical (Dressler 1984, 
and references therein), though Biviano et al. (1995) recently
suggested that this might not be the case for the Coma cluster. In order to 
maintain the luminosity function
unchanged in high-density regions, any ``fading'' of the disk
must then be accompanied by a corresponding brightening of the bulge.
Mergers could be responsible for building up large galactic bulges
in high-density regions. It has been suggested that elliptical galaxies
result from the merging of spiral galaxies (Toomre \&~Toomre 1972;
Toomre 1977; White 1978; 1979; Fall 1979).
Ross (1981) has suggested that galaxies form mainly as stellar disks, and
that galactic bulges are formed by merging, for all galaxy types.
This could explain the fact that the angular momenta of disk and bulge in 
disk galaxies are almost perfectly aligned (Gerhard 1981).
Numerical simulations of galaxy mergers by Mihos \&~Hernquist (1994a, 1994b)
support this model, by showing that mergers trigger infall of material 
toward the center of the system.
This model, if correct, would explain the abundance of {\it both}
S0 and elliptical galaxies in high-density regions. 
Numerical simulations
(Efstathiou \&~Jones 1979; Aarseth \& Fall 1980) have shown that mergers 
of galaxies on highly eccentric orbits result
in the slow-rotating systems, consistent with measurements
of the spin parameter for elliptical galaxies. 

Merging events, however, are not expected to occur inside rich clusters,
where most ellipticals are found. The velocity dispersion in these
regions is quite high, resulting in a significant reduction of the
gravitational cross sections of galaxies. More likely, mergers occur
inside small groups of galaxies where the velocity dispersion is smaller, 
and later these groups assemble into clusters
(see, e.g., Geller \&~Beers 1982). Numerical simulations (Aarseth \&~Fall 1980;
Negroponte \&~White 1983; Noguchi 1988; Barnes 1989; 
Barnes \&~Hernquist 1991; 1992) show that galaxy mergers occur
naturally inside small groups, and that such mergers result in the formation
of spheroidal galaxies with essentially no disk (Barnes \&~Hernquist 1992).
Baugh, Cole, \& Frenk (1996) have used a semi-analytical, Monte Carlo
model to describe galaxy mergers in a standard Cold Dark Matter (CDM) universe.
Their model produces a distribution of D/B
which are consistent with observations, when the values of D/B
are used to ascribe morphological types. Moore et al. (1996) have suggested 
that morphological evolution of spirals actually
occurs inside dense clusters, in spite of the large velocity dispersion.
In their model, called ``galaxy harassment,'' spiral galaxies are disrupted
by the cumulative effect of several high velocity close encounters
with other galaxies.

The various studies of mergers described above consider the merging of
two or more galaxies of comparable size. A completely separate
problem is the merging of a disk galaxy with a satellite galaxy of much
smaller mass. These merging events can modify the structure of the disk,
but the effect is too small to result in actual morphological evolution
(that is, spiral galaxies will remain spiral after ``swallowing'' a 
satellite). Numerical simulations (Quinn \&~Goodman 1986; Quinn,
Hernquist, \& Fullagar 1993;
T\'oth \&~Ostriker 1992) have shown that a merger between a disk galaxy and
a satellite having a mass equal to 1/10 of the mass of the disk results in
a important thickening of the disk, which is ruled out by observations.
However, these simulations ignored the possibility that the satellite
might dissolve significantly before the actual merging takes place. More
recent simulations (Carlberg 1995; Huang 1995) have suggested that the
main effect of these mergers is a tilt of the disk, accompanied by
a transient warp, with no substantial thickening.

There are several problems
with models involving mergers. Elliptical and spiral galaxies
have different globular cluster luminosity functions (Harris 1981).
Since merging events would unlikely affect the structure of globular 
clusters, this result argues against elliptical galaxies being formed
from the merging of spirals, {\it if the number of globular clusters remains
constant during the merging process}. However, Ashman \& Zepf (1992) have
argued that the merging of two galaxies results in the formation of
additional globular clusters. 
Also, dwarf ellipticals presumably {\it do not}
result from mergers, so the continuity of properties for dwarf ellipticals
to regular ellipticals (Sandage 1983) suggests that large elliptical
do not result from mergers either. Merging events would most
likely ruin tight correlations existing among
various parameters for elliptical galaxies, such as color and
luminosity (Bower, Lucey, \& Ellis 1992) and effective radius, central 
velocity dispersion, and mean surface brightness (the ``fundamental plane,''
Djorgovski \& Davis 1987; \hbox{J\o rgensen}, Franx, \& 
\hbox{Kj\ae rgaard} 1996).
Another possible problem is that stars are much more strongly 
concentrated in elliptical galaxies than in spirals (Combes et al. 1995).
However, recent SPH simulations of galaxy mergers (Steinmetz 1995;
Barnes \& Hernquist 1996 and references therein)
show that the merger of two spiral galaxies often results in the formation of
much denser systems, sometimes too dense to be ellipticals galaxies.

\subsubsection{Initial Conditions}

The third class of models comprises models in which the initial conditions 
are primarily responsible for determining the morphological type of galaxies,
with subsequent morphological evolution playing little role or no role
at all. Numerous models have been proposed (see Dressler 1984, and 
references therein), in which the morphological type is determined either by 
the local density, or the local amount of angular momentum.
Such models could successfully explain the observed morphology-density
relation only if galaxies have formed near their present location.
This could be the case in cosmological models which have more
power at large scale than small scale. In such models, clusters would form
first, and then fragment into individual galaxies, in which case galaxies
could actually be located at present near the location were they where formed.
The alternative is that galaxies, at the epoch
of their formation, somehow ``know'' the kind of environment in which
they will be located at the present. This can be achieved if there is
some kind of coupling between the perturbations responsible for forming
the galaxies and the ones responsible for forming the clusters in which
these galaxies end up. 

The problem with these scenarios is that they all invoked cosmological
models that are usually considered ``marginal.'' These models constitute 
interesting alternatives to the more standard CDM
model with Gaussian initial conditions, but there is
at present no strong, absolute evidence favoring such models over the standard
ones. To our knowledge, the most serious alternatives, at present,
to the standard CDM models are the models with Cold + Hot Dark Matter (CHDM),
models with a nonzero cosmological constant, and models with a tilted
power spectrum. None of these models feature
coupling between long- and short-wavelength 
modes in their initial conditions.  
Hence, following Dressler (1984), we will regard these
types of galaxy formation models as ``last resort.'' 

\subsection{Past History of Galaxies}

In order to identify the correct galaxy formation model, we must reconstruct 
the past history of the presently observed galaxies, and in particular we 
need to know the kind of environment in which galaxies were 
located at various 
epochs. We are assuming that galaxies at their formation epoch have no
knowledge of the future environment in which they will end up. We are therefore
rejecting all ``class three'' models, unless galaxies form near their
present location. Hence, if we find that elliptical and S0 galaxies located
in the dense cluster cores were always located in high density environment,
at all epochs up to the galaxy formation epoch (redshifts $z$ of order 3--5),
it would argue in favor of the initial conditions being responsible
for determining the morphological type (class three models), 
and against morphological
evolution. If, to the contrary, many of these elliptical 
and S0 galaxies are found at early time in low density environments, it 
would argue in favor of morphological evolution (class one or two models). 
The goal of this paper is to settle this question.

\section{THE MODELS}

We consider three different cosmological models: an
Einstein-de~Sitter model with $\Omega_0=1$, $\lambda_0=0$,
an open model with $\Omega_0=0.2$, $\lambda_0=0$,
and a low-density flat model with $\Omega_0=0.2$, $\lambda_0=0.8$,
where $\Omega_0$ and $\lambda_0$ are the present values of the
density parameter and cosmological constant, respectively.
We set the present value $H_0$ of the
Hubble constant equal to $50\,\rm km/s/Mpc$ to avoid conflict
between the models and the measurements of globular cluster ages.
With these parameters, the age of the universe $t_0$ is
13.0~Gyr, 16.6~Gyr, and 21.04~Gyr for the Einstein-de~Sitter, open,
and cosmological constant models, respectively.

We assume that the initial fluctuations originate
from a Gaussian random process. The initial density
contrast can then be expressed as a superposition of plane waves
with random phases. Our simulations assume
periodic boundary conditions. This restricts the range of possible values
for the wavenumber~${\bf k}$ to multiples of the fundamental wavenumber
$k_0\equiv2\pi/L_{\rm box}$, where $L_{\rm box}$ is the size of 
the computational volume. The density contrast can then be expressed as
\begin{equation}
\delta({\bf x})=\sum_{\bf k}\delta_{\bf k}
e^{-i{\bf k\cdot x}}\,,
\end{equation}

\noindent
where $\delta_{\bf k}$ is the amplitude of the ${\bf k}$-mode,
and the sum is over all values of ${\bf k}=(l,m,n)k_0$, with
$l$, $m$, $n$ integers.
The requirement that $\delta({\bf x})$ is
real implies~$\delta_{\bf k}=\delta_{-\bf k}^*$. 
The phases of the amplitudes are random, and the norms $|\delta_{\bf k}|$
are related to the power spectrum $P(k)$ by
\begin{equation}
P(k)={V_{\rm box}\over(2\pi)^3}|\delta_{\bf k}|^2\,,
\end{equation}

\noindent where $V_{\rm box}=L_{\rm box}^3$ is the computational volume.
The power spectrum we use can be expressed as 
\begin{equation}
P(k)=AkT(k)^2\,,
\end{equation}

\noindent where $A$ is the amplitude and has dimension of $\rm(length)^4$,
and $T(k)$ is the transfer function. This equation describes an ``untilted''
power spectrum which reduces to the Harrison-Zel'dovich power spectrum
$P(k)\propto k$ at large scale, as $T(k)\rightarrow1$ for $k\rightarrow0$. 
The value of the amplitude is fixed by the value of 
the cosmic microwave background temperature anisotropy, as measured by
COBE (Smoot et al. 1992),
\begin{equation}
A={1\over(2\pi)^3}{6\pi^2\over5}Q_2^2R_H^4\,,
\end{equation}

\noindent where $Q_2$ is the temperature quadrupole anisotropy and $R_H$ is 
the radius of the horizon. For all simulations, we used the value 
$A=8.16\times10^5h^{-4}\rm Mpc^4=1.3056\times10^7\rm Mpc^4$ given by 
Bunn, Scott, \& White (1995) for standard CDM models.
We also use the transfer function given
by Bardeen et al. (1986),
\begin{equation}
T(k)={\cal L}(z){\ln(1+2.34q)\over2.34q}
\big[1+3.89q+(16.1q)^2+(5.46q)^3+(6.71q)^4\big]^{-1/4}\,,
\end{equation}

\noindent
where 
\begin{equation}
q={k\theta^{1/2}\over(\Omega_Xh^2{\rm Mpc^{-1}})}\,,
\end{equation}

\noindent
where $\Omega_X$ is the density parameter of the dark matter (non-baryonic)
component,  
$\theta=1$ for models with 3 flavors of relativistic neutrinos,
and ${\cal L}(z)$ is the linear growth factor between the initial state 
and the present, given by
\begin{equation}
{\cal L}(z)={\delta_+(0)\over\delta_+(z)}\,,
\end{equation}

\noindent where $\delta_+$ is the linear growing mode of the perturbation.
Notice that several different notations are commonly used 
in the literature. Several authors do not include the factor $(2\pi)^3$
in equations~(2) and~(4), and instead include a factor of $(2\pi)^{3/2}$
in equation~(1). Other authors use a redshift-independent transfer 
function, without the ${\cal L}(z)$ factor, and include a factor
of ${\cal L}^2(z)$ in equation~(3).

In all models, we assume that the baryon content of the
universe has a density parameter $\Omega_B=0.0625$.
For the Einstein~de-Sitter model, this gives a density parameter
$\Omega_X=\Omega_0-\Omega_B=0.9375$ for the 
dark matter. For the other two models considered,
$\Omega_0=0.2$, and therefore $\Omega_X$ should be equal to 0.1375, resulting
in a shift of the power spectrum through the relation between $q$ and $k$
given by equation~(6).
Instead, we decided to use the same relationship between $k$ and $q$
for all three models by setting $\Omega_X=0.9375$ in equation~(6), 
thus introducing an inconsistency. Our motivation for doing this is 
the following: Our goal is not to find which model fits the observations of
the present universe better. 
Instead, we want to select cosmological models that 
will bracket the behavior of the large-scale structure formation process.
Using for our initial conditions
a power spectrum that differs among the various models only 
through the model-dependent 
linear growth factor ${\cal L}$ allows us to investigate directly 
the effects of the growth rate and the age of the universe on the evolution of
galaxy clustering. In the same spirit, we are considering open models and 
models with a cosmological constant that are somewhat too extreme to 
agree with observations, which suggests that $\Omega_0$ is more likely
to be somewhere in the range 0.25--0.5 (Ostriker \&~Steinhardt 1995;
Martel, Shapiro, \&~Weinberg 1998, and references therein).
Models with a larger $\Omega_0$ and/or a smaller $\lambda_0$ would
reproduce observations better, but would resemble the Einstein-de~Sitter model
more than the ones we are considering, thus providing less insight on the 
effect of the cosmological parameters on the formation of clusters.
The reader should therefore keep in mind that the power spectrum we are 
using for the open and cosmological constant model is not consistent with 
a standard CDM model, and is chosen only for practical considerations. 

The growing modes $\delta_+(z)$ appearing in equation~(7) are obtained by
solving the linear perturbation equation in the zero-pressure limit.
For the Einstein-de~Sitter model ($\Omega_0=1$, $\lambda_0=0$), 
the growing mode is
\begin{equation}
\delta_+(z)=(1+z)^{-1}\,.
\end{equation}

\noindent
For open models ($\Omega_0<1$, $\lambda_0=0$), the growing mode is
\begin{equation}
\delta_+(z)=1+{3\over x}+3\biggl({1+x\over x^3}\biggr)^{1/2}
\ln\big[(1+x)^{1/2}-x^{1/2}\big]
\end{equation}

\noindent (Peebles 1980), where 
\begin{equation}
x=(\Omega_0^{-1}-1)(1+z)^{-1}\,.
\end{equation}

\noindent Finally, for flat models with a cosmological constant
($\Omega_0+\lambda_0=1$), the growing mode is given by
\begin{equation}
\delta_+(z)=\biggl({1\over y}+1\biggr)^{1/2}
\int_0^y{dw\over w^{1/6}(1+w)^{3/2}}
\end{equation}

\noindent (Martel 1991b), where
\begin{equation}
y={\lambda_0\over\Omega_0}(1+z)^{-3}\,.
\end{equation}

\section{THE CALCULATIONS}

\subsection{The $\bf P^3M$ Algorithm}

All N-body simulations presented in this paper
are done using the {\it Particle-Particle/Particle-Mesh}
(or P$^3$M) algorithm (Hockney \&~Eastwood 1981; Efstathiou \&~Eastwood 1981;
Efstathiou et al. 1985, hereafter EDFW). 
The calculations evolve a system of 
gravitationally interacting particles in a cubic volume with
triply periodic boundary conditions, comoving with Hubble flow.
The forces on particles are computed by solving Poisson equation on a 
$128\times128\times128$ grid using a Fast Fourier Transform method.
The resulting force field represents the Newtonian interaction
between particles down to a separation of a few mesh spacings. At shorter
distances the computed force is significantly smaller than the
physical force. To increase
the dynamical range of the code, the force at short distance 
is corrected by direct 
summation over pairs of particles separated by less than some 
cutoff distance~$r_e$. With the addition of this so-called
{\it short-range correction}, the code accurately reproduces the Newtonian
interaction down to the softening length~$\eta$.
In all calculations, $\eta$ and $r_e$ were set equal to
0.3 and 2.7 mesh spacing, respectively. With these particular values, 
the code has a dynamical range of three orders of magnitude in length (EDFW).
The particular version of P$^3$M we used in this paper uses the so-called
tilde coordinates (Shandarin 1980; Martel \&~Shapiro 1997).
The system is evolved forward in time using a 
second order Runge-Kutta time-integration
scheme with a variable time step.
We define a system of units by setting the mass $M_{\rm sys}$ of the system,
the comoving side $L_{\rm box}$
of the computational volume, and the gravitational constant
$G$ equal to unity.

In all cases, 
the comoving length of the computational volume is $L_{\rm box}=128{\rm Mpc}$
(present length units).
The total mass of the system is
$M_{\rm sys}=3H_0^2\Omega_0L_{\rm box}^3/8\pi G=1.455\times10^{17}
\Omega_0{\rm M}_\odot$.
We use $64^3=262,144$ equal mass particles. The mass per particle 
is therefore $M_{\rm part}=M_{\rm sys}/64^3=5.551\times10^{11}{\rm M}_\odot$ 
for the Einstein-de~Sitter model and $1.110\times10^{11}{\rm M}_\odot$ 
for the other two models.

\subsection{Initial Conditions}

The method we use to set up initial conditions is fairly standard.
We lay down $64^3=262,144$ particles on a uniform cubic lattice, and 
displace them from their initial position in order to
represent the initial density fluctuations.
We then compute the initial peculiar velocities using the
linear perturbation solutions for a pure growing mode,
which are given by equations (8)--(12).

The particle displacements are given by
\begin{equation}
\Delta{\bf x}=-2\sum_{\bf k}
{\delta_{\bf k}{\bf k}\over2\pi k^2}\sin(2\pi{\bf k}\cdot{\bf x}
-\phi_{\bf k})\,,
\end{equation}

\noindent where $\bf x$ the unperturbed position,
$\phi_{\bf k}$ is a random phase between 0 and $2\pi$, and the
sum extends over one half of the $\bf k$-volume (the sine function and the
factor of 2 come from grouping terms in eq.~[1] by pairs with
equal and opposite wavenumbers). In computational units, $k=1$ is the
fundamental mode, whose wavelength is equal the the size $L_{\rm box}$
of the computational volume, and all modes up to the Nyquist frequency
$k=32$ are included.
\footnote{The resulting initial conditions are not truly Gaussian,
because of the discreteness of the sum in equation~(13). This can be 
corrected by choosing the amplitudes $\delta_{\bf k}$ randomly (EDFW),
using a Rayleigh distribution. We decided to ignore this refinement,
since, for the particular combination of particle number and box size 
we are using, these discreteness effects are negligible at scales that
are nonlinear at $z=0$.}

To compute the initial peculiar velocity field, we assume that the initial
time of the calculation is early enough for the perturbation to be
in the linear regime, but late enough so that the linear decaying mode can
be neglected. The initial peculiar velocity of the particles are then 
related to their displacements by
\begin{equation}
{\bf v}_i={\dot\delta_+(z_i)\over\delta_+(z_i)}\Delta{\bf x}\,,
\end{equation}

\noindent where $\Delta{\bf x}$ is computed using equation~(10),
$\delta_+$ is the linear growing mode of the perturbation,
defined by equations~(8)--(12), and $z_i$ is the initial redshift of
the simulations.

\subsection{The Simulations}

We ran 5 simulations for each of the three cosmological models, for
a total of 15 simulation. For each model, the 5~simulations differ
only in the ensemble of random phases used in equation~(13) to generate
the initial particle displacements. To identify these various simulations,
we shall use the following nomenclature: The simulations for the 
Einstein-de~Sitter model ($\Omega_0=1$, $\lambda_0=0$), the
open model ($\Omega_0=0.2$, $\lambda_0=0$), and cosmological constant
model ($\Omega_0=0.2$, $\lambda_0=0.8$) will be called EdSX,
OX, and LX, respectively, where ${\rm X}=1$, 2, 3, 4, 5 identifies
the various runs for each model.
All simulations start at an initial redshift $z_i=24$, and end at $z=0$.

\section{THE PRESENT GALAXY DISTRIBUTIONS AND MORPHOLOGICAL TYPES}

\subsection{The Galaxy Locations}

The P$^3$M algorithm simulates the growth of density fluctuations
resulting in the formation of large-scale structure in an expanding
universe. The only physical interaction present in these simulations is
gravity. Hence, all the hydrodynamical and radiative processes
which certainly play an important role in the galaxy formation process
are ignored. Various authors 
have used P$^3$M codes to simulate galaxy formation, either by 
using a static (Davis et al 1985) or dynamic (Martel 1991a) criterion 
for identifying ``luminous'' particles, by making particles ``stick''
to each others in order to simulate dissipation of kinetic energy by
hydrodynamical processes (Carlberg 1988), or by combining
the P$^3$M algorithm with a hydrodynamical algorithm such as Smoothed
Particles Hydrodynamics
(Evrard 1988). In our simulations, we use a much simpler approach.
We consider the large-scale structure at present ($z=0$) resulting from
the P$^3$M simulations, and design an empirical Monte-Carlo method
for locating galaxies in the computational volume, based on the
constraints that (1) galaxies should be predominantly located in the
densest regions, and (2) the resulting distribution of galaxies
should resemble the observed distribution on the sky.

One possibility consists of using a Monte-Carlo rejection method.
We could generate locations at random inside the computational volume,
and decide whether or not to put a galaxy in these location, based on the 
local density of matter. The likelihood of locating a galaxy in a particular
location should not be a linear function of the local density, however.
Galaxy formation is believed to be biased toward forming galaxies
in high density regions (Kaiser 1984). So in order to use this method,
we would need to know the precise relationship between the matter density and
the likelihood of forming a galaxy. The best currently available theories 
for biased galaxy formation could provide such a relationship,
but using this relationship for locating galaxies would be an overkill.
Biased galaxy formation theories could only provide relationships
that involve the {\it actual} matter distribution in the universe.
We are dealing instead with a {\it simulated} matter distribution,
which is only an approximation of the actual matter distribution.
In particular, CDM models normalized to COBE are known to produce too much
structure on small scales.

Considering these various difficulties, we chose a much simpler
method for locating galaxies. We divide the present
computational volume into $128^3$ cubic cells of size $1{\rm Mpc}^3$, and
compute the matter density $\rho$ at the center of each 
cell, using the same 
mass assignment as in the P$^3$M code. We then choose a particular density
threshold $\rho_{\rm t}$. We locate $N$ galaxies in each cell, where
$N$ is given by
\begin{equation}
N={\rm int}\biggl({\rho\over\rho_{\rm t}}\biggr)\,.
\end{equation}

\noindent The actual location of each galaxy is chosen to be
the center of the cell, plus a random offset of order of the cell size.
This reduces any spurious effect introduced by the use of a grid.
We then experiment with various values of the density
threshold $\rho_{\rm t}$ until the total number of galaxies comes out to be
of order 40000. This gives a number density of 
$\sim0.02\,{\rm galaxies}/{\rm Mpc}^3$.

In Figure~1, we take one simulation for each of the three models,
and plot the location at $z=0$ of the P$^3$M particles
(left panels) and the galaxies (right panel) inside a slice of
size $128\times128\times8\,{\rm Mpc}$. The Einstein-de~Sitter model has too 
much power on small scales, resulting in the formation of very dense clumps.
The cosmological constant model is slightly less evolved, and shows
a large number of average-size clusters that have not yet merged into larger
ones as in the Einstein-de~Sitter model. In this model, the small 
value $\Omega_0=0.2$ of the density parameter results in a small growing rate
of the density fluctuations, but this effect is partly compensated by
the presence of the cosmological constant $\lambda_0$, which increases
the age of the universe and thus allows the fluctuations to grow for a
longer period of time. The open model O1 forms significantly less
structure than the other two.

The galaxies are mostly concentrated in the highest density regions. The
use of a density threshold in equation~(15) approximates quite well the
effect of biased galaxy formation by not locating galaxies in low density
regions. The galaxy distribution for the open model resembles the
observed galaxy distribution. The galaxies in the other 2 models are too 
much clustered. To quantify this point, we compute the 2-point 
correlation function $\xi(r)$ from the simulated galaxy distributions,
for the Einstein-de~Sitter and open models (we omitted the 
cosmological constant model for clarity). The results are shown in Figure~2.
The correlation function for the open model (triangles)
matches the observed power law $\xi(r)=(r/5.4h^{-1}{\rm Mpc})^{-1.77}$ 
(Peebles 1993) (dotted line), for separations $4\,{\rm Mpc}<r<40\,{\rm Mpc}$. 
The correlation function for the Einstein-de~Sitter model (filled circles)
is too large by a factor of 3 over the same range.
This is consistent with results obtained by various authors who have
used more sophisticated methods for generating galaxy distributions 
(see Ostriker 1993, and references therein). Hence, the overclustering
of galaxies in our Einstein-de~Sitter model
should not be regarded as a flaw in our empirical method for locating
galaxies, but rather as a weakness of the CDM model normalized to COBE.
We attribute the excess of correlation at separations $r<4\,\rm Mpc$ for
the open model to the same overmerging problem.

Since there is too much galaxy clustering at present in our
Einstein-de~Sitter model, we can expect that 
earlier time slices will resemble observations better than the present ones.
Using linear perturbation theory, we can approximate the evolution
of the correlation function as $\xi[r/(z+1),z]\approx(1+z)^{-2}\xi(r,z=0)$.
Since the correlation function for the Einstein-de~Sitter model is too
large by a factor of 3 at $z=0$, this 
relation predicts that the $z=3^{1/2}-1\sim0.7$
time slice should match observations better than the present time slice,
which is indeed the case, as shown by the open circles in Figure~2.

One drawback of our empirical scheme for biased galaxy formation is that it 
works ``too well,'' by totally preventing galaxy formation inside voids.
In the real universe, even the deepest voids like Bootes contain some
galaxies, and the existence of these galaxies is significant since is
essentially rules out some cosmological models like Hot Dark Matter.
This limitation of our biasing algorithm is of little consequence for
the argument we present in \S6, however, simply because the actual number
of galaxies located in low density regions is quite small. 

\subsection{The Morphological Types}

As we mentioned in \S1.1,
there is a tight relation between the distribution
of morphological types and the number density of galaxies (Dressler 1984,
and references therein). This morphology-density relation
is reproduced in Figure~3, by the solid curves. By combining
this relation with a Monte-Carlo method, we can ascribe a morphological 
type to each galaxy, as follows. 
We first compute the volume number density of galaxies $\rho_{\rm gal}$
around each galaxy, using 
\begin{equation}
\rho_{\rm gal}={n+1\over4\pi d_n^3/3}\,,
\end{equation}

\noindent where $n$ is a positive integer, and $d_n$ is the distance of the
$n^{\rm th}$ nearest neighboring galaxy. In all cases, we choose $n=12$.
In the case of a spatially uniform distribution of galaxies with
a density $\rho_{\rm uniform}$, this formula gives the correct
answer $\rho_{\rm gal}\approx\rho_{\rm uniform}$ for a galaxy located inside
the distribution, and $\rho_{gal}\approx\rho_{\rm uniform}/2$ for a galaxy
located at the edge of the distribution, since that
galaxy has neighbors on one side only. Notice that Dressler (1980) used
essentially the same technique to compute the surface 
number density of galaxies around each galaxy in his sample.

Once the densities are computed, we compute the fractions 
$f_{\rm Sp}(\rho_{\rm gal})$, $f_{\rm S0}(\rho_{\rm gal})$,
and $f_{\rm Ell}(\rho_{\rm gal})$ 
from the morphology-density relation. We then ascribe a morphological type to
each galaxy by generating a random number $x$ between 0 and 1 (with
uniform probability). The galaxy is a spiral if
$x<f_{\rm Sp}$, a S0 if $f_{\rm Sp}<x<f_{\rm Sp}+f_{\rm S0}$, and
an elliptical if $x>f_{\rm Sp}+f_{\rm S0}$. Table~1 shows the percentages of 
galaxies of each type for each run. Notice that the fluctuations among
different runs within each model are very small. The fluctuations among
different models are larger, and reflect the differences in the amount
of clustering at $z=0$. As we see in Figure~1 (left panels), there is more
clustering in the Einstein-de~Sitter model than in the cosmological constant
model, and significantly more in these two models than in the
open model. This results in a slight excess of elliptical and S0 galaxies
in the Einstein-de~Sitter model compared to the cosmological constant model,
and a bigger excess in these two models compared to the open one. 

Once the morphological types have been assigned, we can compute the
resulting morphology-density relation, and compare it to the
one we were attempting to reproduce. Figure~3 shows the results for the
EdS runs. The error bars indicate the
range of values amongst the 5 different runs for that model, with the
symbols indicating the results obtained by combining all runs together
(this is not the same as the average among the runs, since the various
runs contain different numbers of galaxies). The results reproduce the desired
relations quite well, except at the largest density, where small number
statistics lead to large fluctuations.

\section{TRACING GALAXIES BACK IN TIME}

The P$^3$M algorithm provides us with the distributions of particles
at various intermediate redshifts between the initial redshift
$z=24$ and final redshift
$z=0$. By combining these particle distributions with our
simulated galaxy distributions at present, we can trace galaxies back in time
and reconstruct their trajectory. To do this, we simply find the nearest 
particle $p_i^{(1)}$ of each galaxy $g_i$ at present. Then we ``tie''
the galaxy $g_i$ to that nearest particle. The location of the galaxy
$g_i$ at any redshift $z$ is then given by:
\begin{equation}
{\bf r}[g_i,z]={\bf r}\Big[p_i^{(1)},z\Big]+{\bf r}'\,,
\end{equation}

\noindent where ${\bf r}'$ is a small random offset, which we introduce to
avoid the unfortunate situation of having two galaxies located at the
top of each other because they happen to by tied to the same
particle. This allows us to construct galaxy distributions at any redshift,
and, more importantly, to follow the history of each galaxy as cluster
formation and merging is taking place. Of course, if we trace galaxies
back to redshifts larger than 3--5, we then end up, strictly speaking,
with distributions of {\it protogalaxies}. In Figure~4, we plot the galaxies 
located inside
a slice of comoving thickness $32{\rm Mpc}$ (that is, one quarter of the 
computational volume) at various redshifts, for the run EdS1.

\section{MORPHOLOGICAL EVOLUTION}

\subsection{Elliptical Galaxies}

Knowing the location of each galaxy at various epoch, we can then 
study the local environment in which each galaxy is located, and how
this environment evolves with time.
The basic idea is the following: If elliptical galaxies located in  
the dense cores of clusters 
at $z=0$ were always located in high density environment, 
it will argue against morphological evolution, and
suggest that galaxies formed in such high density
environment form predominantly as ellipticals. If on the contrary, 
many of these elliptical galaxies were located in low density 
environment at, say, $z=3$, it will argue in favor of morphological 
evolution, with these galaxies forming as spiral and later on
becoming elliptical as they find themselves in high density 
environment.

To investigate this question, we compute the number
density of galaxies around 
each galaxy for all 15 runs (3 models with 5 runs for each), at $z=0$
and $z=3$, using the method described in
\S4.2. We are making the null hypothesis that there is no morphological
evolution, hence an elliptical at $z=0$ is also elliptical at $z=3$. We then
sort each list of $\sim40000$ galaxies in increasing order of the local
number density of galaxies.

First, we divide galaxies into low-density environments (L)
and high-density environments (H), both at $z=0$ and $z=3$,
based on the median value of the density at that epoch.
That is, each list contains $\sim20000$ galaxies in low-density 
environments and the same number in high-density
environments. We then divide galaxies into 4 bins according to the type
of environments (L or H) in which they are located at $z=3$ {\it and} $z=0$.
The results are shown in Table~2, where the 
$\rm L\rightarrow L$ bin contains all elliptical galaxies located 
in low-density regions at $z=3$ and $z=0$, the $\rm L\rightarrow H$ bin 
contains the
ones located in low-density regions at $z=3$ and high-density regions at 
$z=0$, and so on. By definition the $\rm L\rightarrow H$ and
$\rm H\rightarrow L$ counts are equal if {\it all} galaxies are considered,
but for now we are only considering elliptical galaxies.

These results show that galaxies are moving through 
environments of different
number densities between $z=3$ and $z=0$. In the Einstein-de~Sitter 
model (runs EdS1 -- EdS5), for instance, only 70\% of the ellipticals
are either $\rm L\rightarrow L$ or $\rm H\rightarrow H$. 
The numbers are very similar among different runs
within each model, showing that these results are statistically significant.
In all cases, the $\rm L\rightarrow H$ count exceeds the
$\rm H\rightarrow L$ count. 
In order to appreciate the significance and implications
of this result, let us consider a simple, probabilistic model
in which the probability that an elliptical galaxy is located in
similar environments at $z=3$ and $z=0$ is $1/2+p$. We obtain the following
relations:
\begin{eqnarray}
{\rm H}_0&=&{\rm H}_3\biggl({1\over2}+p\biggr)
       +{\rm L}_3\biggl({1\over2}-p\biggr)\,,\\
{\rm L}_0&=&{\rm L}_3\biggl({1\over2}+p\biggr)
       +{\rm H}_3\biggl({1\over2}-p\biggr)\,,
\end{eqnarray}

\noindent where ${\rm H}_z$ 
and ${\rm L}_z$, $z=0,3$ are the number of elliptical 
galaxies in high- and low-density environments, respectively, at redshift $z$.
This model has two extreme and opposite limits, which we shall
refer to as the ``no mixing limit'' and the ``complete mixing limit.''
In the no mixing limit, defined by $p=1/2$, each galaxy is 
located at present at or very near
the location (in comoving coordinates) where it was initially formed. Galaxies 
are therefore in identical environments at $z=3$ and $z=0$, and furthermore,
they have the same neighbors. In the complete mixing limit, defined
by $p=0$, all memory
of the location where galaxies were formed has been lost through chaotic
mixing. Any given galaxy can end up at present either in a low- or
high-density environment, with equal probability, no matter 
in which kind of environment it was formed. In this limit, 
${\rm H}_0={\rm L}_0$,
so if the number of galaxies in high- and low-density 
environments at present is actually different,
there is a finite minimum probability $p_{\min}$. We refer to the
case $p=p_{\min}$ as the ``maximum mixing limit.''
We can use this model to analytically compute the 
$\rm L\rightarrow H$ and $\rm H\rightarrow L$ counts and
compare them to the ones given in Table 3, as follows:
We assume that the distribution of galaxies are known at present,
but instead of tracing these galaxies back in time, we now use 
equations~(18) and~(19) to compute the galaxy distributions at $z=3$,
we get
\begin{eqnarray}
{\rm H}_3&=&{{\rm H}_0(1/2+p)-{\rm L}_0(1/2-p)\over2p}\,,\\
{\rm L}_3&=&{{\rm L}_0(1/2+p)-{\rm H}_0(1/2-p)\over2p}\,.
\end{eqnarray}

\noindent by imposing that ${\rm H}_3$ and ${\rm L}_3$ 
are nonnegative, we can solve for the minimum probability,
\begin{equation}
p_{\min}={|{\rm H}_0-{\rm L}_0|\over2({\rm H}_0+{\rm L}_0)}\,.
\end{equation}

\noindent Equations~(20) and~(21) can be solved for any value of $p$
between $p_{\min}$ (maximum mixing limit) and 1/2 (no mixing limit). 
The $\rm L\rightarrow H$ and $\rm H\rightarrow L$ counts are
then given by ${\rm L}_3(1/2-p)$ and ${\rm H}_3(1/2-p)$, respectively.
We plot the results as a function of $p$ in Figure~5, for all three
cosmological models (the values of ${\rm H}_0$ and ${\rm L}_0$ 
used in eqs.~[20] and ~[21]
were obtained by averaging over all five runs within each model).

In all cases, the $\rm L\rightarrow H$ count is lower
than the $\rm H\rightarrow L$ count, for all values of $p$. The cases
$p=p_{\min}$ and $p=1/2$ constitute two extreme 
and opposite limits, no mixing and maximum
mixing, and interestingly {\it these two extreme
limits do not bracket our results}. The reason is that {\it an excess of
$\rm L\rightarrow H$ count over $\rm H\rightarrow L$ count should
not occur ``naturally'' unless there are more galaxies in 
low-density environments
than in high-density environments at $z=3$}, which is clearly
not the case for ellipticals in our simulations. 

This seemingly absurd 
result is based on the assumption that there is no morphological 
evolution. If we relax this assumption, we can solve the 
``$\rm L\rightarrow H$ excess'' problem. 
If some elliptical galaxies located in high-density environments at present 
were
actually formed as spiral galaxies in low-density environments, and eventually
became ellipticals as they found themselves in higher-density
environments at later time, then we are overestimating the 
$\rm L\rightarrow H$
count by ignoring morphological evolution. In order to bring the
$\rm L\rightarrow H$ count down to the value of the $\rm H\rightarrow L$
count or lower, we must speculate that at least 1/4 of the elliptical galaxies
in the $\rm L\rightarrow H$ bin were formed as spiral galaxies, and underwent
morphological evolution between $z=3$ and $z=0$ that transformed them
into elliptical galaxies.

The same probabilistic model can be applied to other morphological types.
Since the {\it total} $\rm L\rightarrow H$ and $\rm H\rightarrow L$ counts
must be equal by definition, at least one type of galaxy must have
an excess of $\rm H\rightarrow L$ over $\rm L\rightarrow H$ to compensate
for the ellipticals. This is indeed the case for the spirals. Applying the
same probabilistic model to the spirals, we would find that, for all
allowed values of $p$, the ``natural'' tendency for spirals is to move from
low density regions to high density regions, simply because there are more
spirals in low density regions to start with. We can solve this 
``$\rm H\rightarrow L$ excess'' among spirals by assuming that some spiral
galaxies turned into ellipticals as they moved into high density regions, 
leading to an underestimate of the $\rm L\rightarrow H$ count.

\subsection{All Morphological Types}

In this subsection, we consider galaxies of all types 
(not only ellipticals) that have formed in
low-density environments. The results are shown in Table~3,
where the numbers in parentheses are the percentages for each type.
We are still making the null hypothesis of no morphological evolution.
The percentages are different for the
$\rm L\rightarrow L$ and $\rm L\rightarrow H$ bins, which is of course
totally absurd: It implies that, somehow, the galaxy formation process
is able to ``distinguish'' a low density environment at $z=3$ that will 
remain low-density at all times from one that will eventually become
high-density. Since we assume 
there is no ``fortune teller'' at $z=3$ that can ``tell'' 
the galaxy formation process what will happen in
the future, thus excluding class~3 models,
we must conclude that morphological evolution is present.
We can reconcile the numbers
presented in Table~3 by assuming that spiral galaxies evolve either into
S0 or ellipticals galaxies. For instance, we can reconcile the percentages
for the EdS1 run by ``transferring'' 534 galaxies from the 
$\rm L\rightarrow H$ S0 bin to the $\rm L\rightarrow H$ Spiral bin,
and 159 galaxies from the $\rm L\rightarrow H$ Elliptical bin 
to the $\rm L\rightarrow H$ Spiral bin. The percentages would then be
the same as for the $\rm L\rightarrow L$ bins.
This would imply that 21\% of these
S0 galaxies (534 out of 2565) and 17\% of these elliptical galaxies 
(159 out of 924) were formed as spiral galaxies and
underwent morphological evolution at a later time.

One possible problem with this interpretation of the results is our
definition of low-density and high-density environments. 
At $z=3$, for the Einstein-de~Sitter model, the number density of galaxies
around each galaxy varies from $1.3\times10^3$ to $2.6\times10^7$ 
per unit computational volume with the
median being $2\times10^5$
(the comoving number density is obtained by dividing
these numbers by $[128\,{\rm Mpc}]^3$; the physical number density is 
obtained by dividing
these numbers by $[128(1+z)^{-1}\,{\rm Mpc}]^3$). 
Hence, the number density in ``low density environments''
defined as the bottom half of the distribution, varies over 2 orders
of magnitude. Then we could argue that the galaxies ending up in low-
and high-density environments at $z=0$ come from different 
``parts'' of the low-density environments at $z=3$. 

To solve this problem, we define a ``very low density'' (VL) 
environment, comprising all galaxies located in the bottom 1/20 of the 
number density
distribution, that is, the $\sim2000$ of these $\sim40000$ 
galaxies that are located
in the least dense environments. The number
density at $z=3$ around these galaxies
varies from $1.3\times10^3$ to $2.8\times10^4$, 
but if we ignore a small number of galaxies 
in {\it extremely} low density environments (about 50), the range becomes 
$5.0\times10^3-2.8\times10^4$. 
The physical conditions for galaxy formation in these
regions should be quite uniform, hence the percentages of spirals,
S0's, and ellipticals
should be essentially the same everywhere within these regions. We then look
at the location of these galaxies at $z=0$. The results are shown
in Table~4.

Again, these percentages are different depending on whether
the galaxies end up in low- or high-density environments at $z=0$ (the
results for the open models are statistically insignificant, because
only a few galaxies ended up in high-density environments). Since the galaxy 
formation process cannot predict which galaxies will end up in high or
low-density environments at $z=0$, we are forced to reject the null 
hypothesis of no morphological evolution.

Again, we can reconcile the numbers
presented in Table~4 by ``transferring'' galaxies from the
$\rm VL\rightarrow H$, S0 and Elliptical bins to the 
$\rm VL\rightarrow H$ Spiral bin. For the run EdS1,
transferring 23 galaxies from the 
$\rm VL\rightarrow H$ S0 bin to the $\rm VL\rightarrow H$ Spiral bin,
and 5 galaxies from the $\rm VL\rightarrow H$ Elliptical bin 
to the $\rm VL\rightarrow H$ Spiral bin
would make the percentages the same as for the $\rm VL\rightarrow L$ bins.
Hence, 28\% (23 out of 82) of these
S0 galaxies and 17\% (5 out of 30) of these elliptical galaxies formed as 
spiral galaxies. Notice the similarity of these percentages with the
ones computed from Table~3. 

These numbers are smaller
if we assume that morphological evolution involves galaxy collision and
merging. In the simplest case, morphological evolution transforms
2 interacting spiral galaxies into one S0 or elliptical galaxy.
Hence, for each S0 or elliptical galaxy we ``remove''
from their $\rm VL\rightarrow H$ bin, we need to add 2 spiral
galaxies, instead of only one, to the Spiral bin. In this case, to reconcile
the percentages for the EdS1 run, we need to remove 18 S0 galaxies
and 2 elliptical galaxies, thus adding 40 spiral galaxies
($2\times[18+2]$). The fractions of S0 and elliptical galaxies that were
formed by mergers then becomes 22\% (18 out of 82) and 7\% (2 out of 30).

\section{THE EVOLUTION OF CLUSTERING}

The results of the previous section suggest that some elliptical and S0
galaxies were formed as spiral galaxies and underwent morphological
evolution at some epoch between redshifts of $z=3$ and $z=0$.
Assuming that the morphological evolution process is triggered by an increase
in the galaxy number density resulting from the
formation and merging of clusters, we can attempt to estimate the epoch of 
galaxy evolution by monitoring the evolution of the number density of galaxies
around the galaxies that might have undergone 
such morphological evolution.

In Figure~6, we plot, as a function of redshift, the number density $n$ of
galaxies (in $\rm galaxies/Mpc^3$)
around each elliptical galaxy located in the 
$\rm VL\rightarrow H$ bin of Table~4
(runs O2, O4, and O5 do not contain any such galaxy, and are
therefore omitted). The number of curves in 
each panel can be read from the last column of Table~4.
All panels show a similar
pattern. Initially, the number density decreases with time, indicating that
these regions are still expanding 
(though slower than Hubble flow). The number densities
reach a minimum at epochs between $z=0.4$ and $z=0.8$, indicating that
the regions surrounding these galaxies have finally turned back and started
to recollapse. The number densities then increase by 2 to 3 orders of magnitude
between the turnaround epoch and the present, except for the open model
(runs O1 and O3) for which the density increase is smaller than one order
of magnitude. Some galaxies in the EdS runs (and also one in the L3 run)
follow a different history, with the number density starting to rise at
$z\sim1$, then dropping and rising again. The initial increase is caused by
the formation of a dense cluster at $z\sim1$, resulting from the
collapse of a particularly large density fluctuation. The subsequent
drop in number density is caused by the tidal disruption of that cluster by 
more massive clusters formed at later epochs. These cases constitute a 
minority.

These plots indicate that the morphological evolution process, if real, 
most probably takes place at redshifts smaller than $z=0.6$, after the
number density of galaxies has started to increase. Furthermore, the number
densities reach the same value they had at $z=3$ at a redshift of
order $z\sim0.2$. It is tempting to argue that, for morphological 
evolution to occur, the number density has to get larger than it was
at the galaxy formation epoch, and therefore it must occur between $z=0.2$
and $z=0$. This argument is not valid because it assumes
that the morphological evolution process depends {\it directly}
on the number density of galaxies, which is presumably not the case.
If morphological evolution results from galactic collisions or tidal stripping,
then the likelihood for this process actually happening will depend upon
the probability of having close encounters between galaxies, which is larger
in regions of high number density. However, for the same number density, the
likelihood of having close encounters between galaxies is much larger
at $z=0.2$ than at $z=3$. Not only are galaxies more clustered at $z=0.2$
(see Fig.~4 for a good illustration of that), but in addition the galaxies,
overall, are moving apart from one another at $z=3$, whereas they
are approaching each other at $z=0.2$. Hence, we cannot rule out the 
possibility that morphological evolution takes place between redshifts
$z=0.8$ and $z=0.2$ on the basis that the number densities at these epochs
are smaller than they are at $z=3$.

Several authors have claimed that
merging events were more frequent in the past, based either
on observations (see, e.g. Carlberg 1995 and references therein)
or analytical arguments (Toomre 1977; Aarseth \&~Fall 1980).
These results do not contradict our claim that morphological evolution
does not occur at redshifts $z>0.8$, simply because
we are focusing our attention to {\it very low density regions}.
In particular, the analytical arguments aforementioned assume that
merging involves pairs of galaxies which are already on bound
orbits, which is clearly not the case in the regions we are
considering, which are still dominated by an overall expansion at $z=3$.
Also, galaxy merging is only one of many physical processes that could possibly
result in morphological evolution. In this section, we are making no
assumption on the nature of the actual physical process involved.
We are merely arguing that morphological evolution in 
very low density regions does not occur until $z\sim0.8$, 
simply because at earlier
time all galaxies in these regions are moving away from one another.
Notice that this result is based on galaxies located in VL regions at $z=3$.
Higher density regions would turn back at larger redshift. 

\section{DISCUSSION OF THE METHOD}

In this section, we review and discuss the strengths and weaknesses 
of our numerical simulations, and the interpretation of the results.

\subsection{Weak Points}

The weakest point of this entire work is certainly the cosmological
simulations themselves. The Einstein-de~Sitter model with 
CDM spectrum normalized to COBE is known to produce
too much structure at small scale. This is reflected in the two-point
correlation function which is too large by a factor of 3 in the range
of 1 -- 10~Mpc. Only in the open model does the distribution of galaxies
actually resemble the present universe. Since the 2-point correlation
function
evolves roughly as $a(t)^2\propto(1+z)^{-2}$ in the linear regime, we can
estimate that $z\sim0.7$ (that is,
$3^{1/2}-1$) time-slices for these models would 
be a better representation of the actual present universe. We looked at these
time-slices, and they indeed resemble the present universe more closely
than the $z=0$ time-slice. 

It is difficult to estimate the consequences of this excess of 
small-scale structure. We argue that the effect is not so significant,
and does not affect our conclusion. The main point is that we get
consistent results among all three models, including the open model
which {\it does} reproduce the present universe fairly well.
Also, it is hard to see how the excess of structure formation in
the Einstein-de~Sitter model and the cosmological constant model
could possibly affect the conclusion. Cluster merging happens continuously
in CDM models, all the way to the present. The excess of power simply
increases the amount of merging taking place between $z=0$ and $z=3$. 
We divide the regions in which galaxies end up at $z=0$ into
low density and high density environment, without taking into
account how high the number density gets inside these regions
of high density. Hence, late cluster mergers are unlikely to have
a strong effect on the results shown in Tables~2--4. As long as we are not
interested in galaxies located in ``very high density environments'' at
$z=0$, the excess of structures at small scale is probably unimportant.

We traced the motion of galaxies back in time by following the motion of
the nearby dark matter particles. This assumes that the velocity field 
of galaxies and dark matter are the same. This assumption is certainly
valid at early times. However, numerical simulations (Carlberg 1994) have
shown the existence of a velocity bias between galaxies and dark matter
inside clusters of galaxies. This is the result of an evolutionary process
taking place inside the clusters. Hence, our method for tracing galaxies
back in time might be partly flawed if this velocity bias is real.

Finally, our biasing scheme for galaxy formation is quite crude. This
is certainly an aspect of the algorithm that could use some improvement.
Unfortunately, not much can be done until the cosmological models themselves
are improved. No biasing scheme will ever be satisfactory as long as the 
cosmological simulations produce too much structure at small scale.

\subsection{Strong Points}

The strongest point of this entire work is that the conclusions do not
depend on the details of the initial galaxy formation and 
morphological evolution processes. The only assumptions we make
concerning the initial galaxy formation process
are that (1) it takes place before $z=3$, and (2) it has no ``knowledge''
of the future. As for the morphological 
evolution process, the only assumptions we
make are that it converts spiral galaxies into S0 and elliptical galaxies,
but not the other way around, and that it takes place in high 
density environments. The detailed physical processes involved in the initial
galaxy formation and 
morphological evolution processes are irrelevant to this work,
and this only makes our results more robust.

The second strongest point is the consistency of our results, first among
different simulations for a same cosmological
model, and then among the various
models. The percentages shown in Tables~3 and~4 
(with the exception of the Open model on Table~4)
have error bars much smaller
than the differences among these various percentages, which is what our
argument is based on. Also, the fact that all three cosmological
models show a trend toward morphological evolution strongly suggests that this
effect is real.

There is a potential problem with the technique we use for tracing galaxies
back in time. If dense clusters form by assembling matter taken from distant
regions of the universe
(which might be the case when cluster mergers are involved),
then our approach of tying each galaxy to the nearest dark matter particle
becomes ambiguous. A given galaxy might have formed in any of these 
distant regions, and by following the trajectory of the nearest dark matter
particle, we are ``forcing'' that galaxy to have formed in one particular
region, when it could actually have formed in another one.

To estimate the importance of this effect, we go back to the run EdS1,
\footnote{We choose a run from the EdS model because the structures
are more evolved in this model than in the other ones, with more cluster
merging happening at late time. Hence, the effect we are trying to 
measure is likely to be more important in this model.} and recompute
the trajectories of the galaxies, except that we replace
$p_i^{(1)}$, the nearest dark matter particle to each galaxy, by
$p_i^{(2)}$, the second-nearest particle, in equation~(17). 
We label this new calculation EdS1$^*$. In Figure~7a, we plot the
$x$-coordinate, in computational units, of the galaxies at $z=3$ for
the EdS1$^*$ run, versus the same coordinates for the EdS1 run.
Even though there is some scatter, most galaxies are located near the
diagonal, indicating that the differences between the two runs are
small for most galaxies (the concentrations of galaxies in the upper
left and lower right corners of the figure are an artifact of the
periodic boundary conditions). Plots of the $y$- and $z$-coordinates
are similar. For brevity, we are omitting them in this paper.

Figure~7b shows an histogram of the 3-dimensional separation, in computational
units, between each galaxy at $z=3$ in the EdS1 run and its counterpart
in the EdS1$^*$ run. More than 1/3 of the galaxies are located 
in the first bin, having separations less than 1/40 [corresponding to 
a physical separation of
$128\,{\rm Mpc}(z+1)^{-1}/40=800\,\rm kpc$], and the first seven bins contain
92\% of the galaxies. Hence, only a few galaxies end up in significantly
different regions when we track the second-nearest particle
instead of the first one.

Using the galaxy locations
for the run EdS1$^*$, we perform the same analysis 
as for the other runs. The results are given in the last line of Table~4.
The numbers for the run EdS1$^*$ are remarkably similar to the ones
for the EdS1 run. The most important difference is in the fraction
of elliptical galaxies in the $\rm VL\rightarrow H$ bin, which is
30\% in one case and 42\% in the other. But actually, the
EdS1$^*$ run is closer to the average amongst EdS runs than the EdS1 run is. 
Therefore, following the trajectory of the second-nearest particle
instead of the nearest one does not affect our final conclusion in any way.

Finally, our method is based on comparing the number of galaxies in
various bins (for instance, the number of elliptical galaxies in
$\rm VL\longrightarrow L$ and $\rm VL\longrightarrow H$ bins). The fact
that our conclusions are based, not on the galaxy counts themselves, but on
comparisons between counts, offsets some drawbacks of the cosmological models.
The CDM model normalized to COBE produces an excess of dense regions, and as
a result our simulations contain more elliptical and S0 galaxies than the 
real universe. If we had fewer elliptical galaxies in our models, the
counts in the $\rm VL\longrightarrow L$ and $\rm VL\longrightarrow H$
bin would most likely be reduced {\it by the same factor}, and our 
conclusion would be the same. Our method uses elliptical galaxies as 
mass tracers, and having an excess of such tracers simply improves 
statistics.

\section{CONCLUSION AND PROSPECTS}

We conclude that a small but non-negligible fraction (of order 10\%--20\%)
of the S0 and elliptical galaxies we observe today in the dense parts 
of clusters were not formed as S0's and ellipticals, but rather as spiral
galaxies,
and underwent morphological evolution between $z=3$ and $z=0$, presumably
during cluster formation and merging. Since the fraction of galaxies involved
in morphological evolution is neither 0\% nor 100\%, initial conditions
and morphological evolution processes must {\it both} play an important role
in determining the morphological type of galaxies.

Our simulations predict that
the proportion of spiral galaxies should increase form the present
observed value of $\sim50\%$ to larger values as one looks back in time,
that is, at larger redshifts. However, they cannot predict
at what redshift this effect would manifest itself, and consequently 
we cannot predict the shape of the morphology-density relation
at high redshift. To make a theoretical prediction, 
we need first to understand the details of the morphological evolution 
process. Also, the epoch of galaxy formation most certainly depends
upon the cosmological model, so before we can make quantitative predictions,
we first need to settle the question of which cosmological model properly
describes the formation of large-scale structures in the universe.  

However, a large amount of observational evidence supporting the
existence of morphological evolution in dense environments at 
redshifts $z<0.5$ has been accumulated in recent years.
Butcher \& Oemler (1978, 1984)
discovered a large excess of blue objects in clusters located at redshift
$z\gtrsim0.4$. Subsequent ground-based observations
(Dressler \&~Gunn 1982, 1983; Couch et al. 1983; Couch \& Newell 1984; 
Dressler, Gunn, \&~Schneider 1985; Ellis et al. 1985; Lavery \&~Henry 1986; 
Henry \&~Lavery 1987; Couch \&~Sharples 1987; MacLaren, Ellis, \&~Couch 1988; 
Soucail et al. 1988; Arag\'on-Salamanca, Ellis, \&~Sharples~1991;
Arag\'on-Salamanca et al. 1993) have
shown that this ``Butcher-Oemler effect'' results from
short-lived bursts of star formation affecting a subset of the cluster members.
These starbursts could be triggered by the ram pressure of the intracluster 
gas when a galaxy first enters the cluster, by violent interaction
between galaxies, or by mergers, 
(see Bothun \&~Dressler 1986, and references therein;
Oemler 1992, and references therein; Mihos \&~Hernquist 1994a, 1994b).
Recent Hubble Space Telescope observations of high redshifts clusters
$z\sim0.3-0.5$
revealed that the blue starburst objects are low luminosity spiral galaxies,
with as many as $\sim50\%$ of them being disturbed by what appears to be 
either tidal disruption or merging (Dressler et al. 1994a, 1994b; 
Couch et al. 1994; Barger et al. 1995). The galaxy populations of these 
clusters differ significantly form the ones of nearby clusters,
and resemble the ones seen in the nearby small groups and field. 

The difference between the galaxy populations of high-redshift and
low-redshift clusters and the importance of dynamical interaction in 
high-redshift clusters compared to low-redshift ones
provide strong evidence that morphological evolution
has occurred inside rich clusters. Studies of galaxy
populations in the field (Colless et al. 1990; Griffiths et al. 1994;
Mobasher et al. 1996) 
and in small groups (Allington-Smith et al. 1993), reveal that no
significant morphological evolution has occurred
in these environments between redshift $z=0.5$ and the present, at least 
among luminous galaxies.
(Driver et al. [1995], however, found an excess of {\it faint}
late type galaxies in the field.) These results rule out any model 
in which the morphological evolution of a galaxy is driven by an
internal physical process. The morphological evolution process depends
upon the richness of the environment, and thus results in a steepening of the
morphology-density relation with time.
 
The most recent studies (Dressler \& Smail 1996; Smail et al. 1997;
Dressler et al. 1997, and references therein)
of high-redshift clusters,
which include 10 rich clusters ($0.36<z<0.57$) comprising 1857
galaxies, show that the excess of spiral galaxies in high-redshift clusters
is compensated by an underabundance of S0 galaxies, but not ellipticals.
This implies that if morphological evolution is responsible for
forming both some S0 and some elliptical galaxies, 
as our numerical simulations
suggest, then the process of converting spirals galaxies into ellipticals
must have occurred {\it before} $z=0.57$. 
This hypothesis constitutes an observational challenge, since testing it
requires observations of even more distant clusters, in the range
$z\sim0.5-0.8$, and a theoretical challenge as well, finding
a model that explains why morphological evolution produces S0 and
elliptical galaxies at different epochs.

All these observational results are consistent with our conclusion that
a fraction of the elliptical and S0 galaxies result from morphological
evolution processes taking place between redshifts of order unity and the 
present. The observations and our numerical simulations
both indicate that the correct galaxy formation model ought to be a 
``class two'' model, in which both initial conditions and 
morphological evolution play an important role. Finding the correct 
galaxy formation model
will most likely require a better understanding of the physical 
processes involved and the cosmological context in which they are taking
place, as well as observations and determination
of morphological types in clusters beyond redshift $z\sim0.5$.

\acknowledgments

This work benefited from stimulating discussions with
Alan Dressler, Inger J\o rgensen, George Lake, and Paul Shapiro.
We are pleased to acknowledge the support of NASA Grant NAG5-2785,
NSF Grants PHY93~10083 and ASC~9504046,
the University of Texas High Performance Computing Facility
through the office of the vice president for research,
and Cray Research.

%

%

\clearpage
\begin{deluxetable}{cccc}
\tablecaption{MORPHOLOGICAL TYPES AT $z=0$. \label{tbl-1}}
\tablewidth{0pt}
\tablehead{
\colhead{Run} & \colhead{Ellipticals} & \colhead{S0's} & \colhead{Spirals}}
\startdata
EdS1 & 14.5 & 38.5 & 47.0 \nl
EdS2 & 14.7 & 38.7 & 46.6 \nl
EdS3 & 14.6 & 38.6 & 46.8 \nl
EdS4 & 14.8 & 38.6 & 46.6 \nl
EdS5 & 14.8 & 38.7 & 46.5 \nl
\tableline
O1   & 13.3 & 35.9 & 50.9 \nl
O2   & 13.4 & 35.6 & 51.0 \nl
O3   & 13.3 & 35.0 & 51.7 \nl
O4   & 13.4 & 35.7 & 50.9 \nl
O5   & 13.1 & 35.1 & 51.8 \nl
\tableline
L1   & 14.4 & 37.8 & 47.8 \nl
L2   & 14.7 & 38.0 & 47.3 \nl
L3   & 14.3 & 37.7 & 48.0 \nl
L4   & 14.3 & 37.8 & 47.9 \nl
L5   & 14.6 & 38.1 & 47.3 \nl
\enddata
\end{deluxetable}

\clearpage
\begin{deluxetable}{ccccc}
\tablecaption{TRANSFER OF LOCATION FOR ELLIPTICAL GALAXIES FROM $z=3$ TO $z=0$
\label{tbl-2}}
\tablewidth{0pt}
\tablehead{
\colhead{Run} & \colhead{${\rm L}\rightarrow{\rm L}$} & 
\colhead{${\rm L}\rightarrow{\rm H}$} &
\colhead{${\rm H}\rightarrow{\rm L}$} & \colhead{${\rm H}\rightarrow{\rm H}$}}
\startdata
EdS1 & 1823 &  924 &  785 & 2440 \nl 
EdS2 & 1758 &  976 &  820 & 2434 \nl
EdS3 & 1704 & 1048 &  840 & 2355 \nl
EdS4 & 1780 & 1079 &  845 & 2467 \nl
EdS5 & 1769 & 1076 &  817 & 2517 \nl
\tableline
O1   & 1686 &  754 &  552 & 2412 \nl
O2   & 1689 &  745 &  600 & 2391 \nl
O3   & 1634 &  718 &  623 & 2423 \nl
O4   & 1675 &  759 &  630 & 2446 \nl
O5   & 1618 &  764 &  640 & 2217 \nl
\tableline
L1   & 1834 &  945 &  762 & 2441 \nl
L2   & 1826 &  995 &  752 & 2534 \nl
L3   & 1772 &  929 &  781 & 2514 \nl
L4   & 1790 &  947 &  759 & 2426 \nl
L5   & 1774 & 1028 &  774 & 2502 \nl
\enddata
\end{deluxetable}

\clearpage
\begin{deluxetable}{ccccccc}
\tablecaption{TRANSFER OF LOCATION FOR ALL GALAXIES IN LOW DENSITY 
REGIONS, FROM $z=3$ TO $z=0$ \label{tbl-3}}
\tablewidth{0pt}
\tablehead{
\colhead{Run} 
& \multicolumn{3}{c}{$\displaystyle\strut{\rm L}\rightarrow{\rm L}\over{
\displaystyle\strut\rm Spirals\qquad\quad S0's\qquad\quad Ellipticals}$} 
& \multicolumn{3}{c}{$\displaystyle\strut{\rm L}\rightarrow{\rm H}\over{
\displaystyle\strut\rm Spirals\qquad\quad S0's\qquad\quad Ellipticals}$}}
\startdata
EdS1 & 7813 (53.9) & 4863 (33.5) & 1823 (12.6) 
     & 2576 (42.5) & 2565 (42.3) &  924 (15.2) \nl
EdS2 & 7596 (52.9) & 4994 (34.8) & 1758 (12.3) 
     & 2518 (41.4) & 2584 (42.5) &  976 (16.1) \nl
EdS3 & 7487 (53.2) & 4879 (34.7) & 1704 (12.1) 
     & 2692 (42.6) & 2586 (40.9) & 1048 (16.6) \nl
EdS4 & 7671 (53.0) & 5010 (34.6) & 1760 (12.3) 
     & 1686 (42.2) & 2603 (40.9) & 1079 (16.9) \nl
EdS5 & 7699 (53.0) & 5067 (34.9) & 1769 (12.2) 
     & 2699 (41.4) & 2743 (42.1) & 1076 (16.5) \nl
\tableline
O1   & 8963 (58.6) & 4654 (30.4) & 1686 (11.0) 
     & 2322 (45.7) & 2000 (39.4) &  754 (14.9) \nl
O2   & 9018 (59.4) & 4480 (29.5) & 1689 (11.1) 
     & 2437 (46.4) & 2072 (39.4) &  746 (14.2) \nl
O3   & 9101 (60.0) & 4429 (29.2) & 1634 (10.8) 
     & 2433 (47.8) & 1948 (38.1) &  718 (14.1) \nl
O4   & 9057 (58.9) & 4655 (30.3) & 1675 (10.9) 
     & 2423 (46.4) & 2035 (39.0) &  760 (14.6) \nl
O5   & 8803 (59.1) & 4473 (30.0) & 1618 (10.9) 
     & 2421 (46.8) & 1987 (38.4) &  764 (14.8) \nl
\tableline
L1   & 8055 (54.7) & 4842 (32.9) & 1834 (12.4) 
     & 2614 (42.9) & 2528 (41.5) &  945 (15.5) \nl
L2   & 7956 (54.2) & 4901 (33.4) & 1826 (12.4) 
     & 2586 (42.6) & 2493 (41.0) &  995 (16.4) \nl
L3   & 8139 (55.1) & 4906 (33.4) & 1723 (11.7) 
     & 2589 (43.3) & 2462 (41.2) &  929 (15.5) \nl
L4   & 8083 (54.8) & 4872 (33.0) & 1790 (12.1) 
     & 2584 (43.4) & 2424 (40.7) &  947 (15.9) \nl
L5   & 8136 (55.2) & 4830 (32.8) & 1774 (12.0) 
     & 2595 (42.5) & 2487 (40.7) & 1028 (16.8) \nl
\enddata
\end{deluxetable}

\clearpage
\begin{deluxetable}{ccccccc}
\tablecaption{TRANSFER OF LOCATION FOR ALL GALAXIES IN VERY LOW DENSITY 
REGIONS, FROM $z=3$ TO $z=0$ \label{tbl-4}}
\tablewidth{0pt}
\tablehead{
\colhead{Run} 
& \multicolumn{3}{c}{$\displaystyle\strut{\rm VL}\rightarrow{\rm L}\over{
\displaystyle\strut\rm Spirals\qquad\quad S0's\qquad\quad Ellipticals}$} 
& \multicolumn{3}{c}{$\displaystyle\strut{\rm VL}\rightarrow{\rm H}\over{
\displaystyle\strut\rm Spirals\qquad\quad S0's\qquad\quad Ellipticals}$}}
\startdata
EdS1 & 1124 (61.1) &  499 (27.1) &  217 (11.8) 
     &  104 (48.1) &   82 (38.0) &   30 (13.9) \nl
EdS2 & 1076 (60.2) &  489 (27.4) &  221 (12.4) 
     &  113 (44.1) &  107 (41.8) &   36 (14.1) \nl
EdS3 & 1018 (58.6) &  514 (29.6) &  205 (11.8) 
     &  129 (42.7) &  126 (41.7) &   47 (15.6) \nl
EdS4 & 1066 (58.3) &  562 (30.7) &  202 (11.0) 
     &   99 (39.3) &  101 (40.1) &   52 (20.6) \nl
EdS5 & 1133 (61.7) &  529 (28.8) &  174 ( 9.5) 
     &  106 (39.4) &  122 (45.4) &   41 (15.2) \nl
\tableline
O1   & 1443 (71.2) &  391 (19.3) &  193 ( 9.5) 
     &    2 (20.0) &    4 (40.0) &    4 (40.0) \nl
O2   & 1491 (73.1) &  367 (18.0) &  183 ( 9.0) 
     &    2 (66.7) &    1 (33.3) &    0 ( 0.0) \nl
O3   & 1429 (70.7) &  367 (18.2) &  225 (11.1) 
     &    1 (25.0) &    2 (50.0) &    1 (25.0) \nl
O4   & 1474 (71.8) &  374 (18.2) &  206 (10.0) 
     &    5 (83.3) &    1 (16.7) &    0 ( 0.0) \nl
O5   & 1435 (72.0) &  373 (18.7) &  184 ( 9.2) 
     &    9 (64.3) &    5 (35.7) &    0 ( 0.0) \nl
\tableline
L1   & 1214 (64.3) &  459 (24.3) &  215 (11.4) 
     &   91 (47.2) &   68 (35.2) &   34 (17.6) \nl
L2   & 1246 (64.5) &  478 (24.8) &  207 (10.7) 
     &   71 (49.3) &   49 (34.0) &   24 (16.7) \nl
L3   & 1238 (65.3) &  477 (25.2) &  180 ( 9.5) 
     &   72 (40.2) &   85 (47.5) &   22 (12.3) \nl
L4   & 1251 (65.2) &  470 (24.5) &  198 (10.3) 
     &   71 (47.0) &   50 (33.1) &   30 (19.9) \nl
L5   & 1272 (65.5) &  455 (23.4) &  215 (10.3) 
     &   62 (43.4) &   63 (44.1) &   18 (12.6) \nl
\tableline
EdS1$^*$ & 1119 (62.1) & 489 (27.2) & 193 (10.7) 
         &  105 (41.2) & 108 (42.4) &  42 (16.5) \nl
\enddata
\end{deluxetable}

%

\clearpage
\begin{center}
Figure Captions
\end{center}

\figcaption[]
{(a) $x-y$ projection of the final positions of the dark
            matter particles for the run EdS1, with $\Omega_0=1$,
            $\lambda_0=0$. (b) $x-y$ projection of the final positions
            of the galaxies for the run EdS1. (c) and (d): same as
            (a) and (b) for the run O1, with $\Omega_0=0.2$, 
            $\lambda_0=0$. (e) and (f): same as
            (a) and (b) for the run L1, with $\Omega_0=0.2$, 
            $\lambda_0=0.8$. On all panels, only 1/32 of the
            computational volume is plotted. \label{fig1}}

\figcaption[]
{Galaxy 2-point correlation function versus separation,
            for the Einstein-de~Sitter model (filled circles) and
            the open model (triangles), both at $z=0$, and
            for the Einstein-de~Sitter model at $z=0.7$ (open circles).
            The dashed line indicates
            the observed correlation function. \label{fig2}}

\figcaption[]
{Population distributions of the various morphological types,
            vs galaxy number density. Solid curves show the inferred 
            relation, based on the observed morphological type-surface
            number density of Dressler (1980). Symbols show the numerically
            generated distributions for all 5 calculations with $\Omega_0=1$,
            $\lambda_0=0$ combined. Error bars show the range
            of values amongst the various calculations. Crosses: Spiral
            galaxies; Filled circles: S0 galaxies; Open circles: Elliptical
            galaxies; The number density $\rho$ is in galaxies/Mpc$^3$.
\label{fig3}}

\figcaption[]
{Galaxy distributions at various redshifts for the run EdS1.
            Only 1/8 of the computational volume is plotted.\label{fig4}}

\figcaption[]
{$\rm L\rightarrow H$ and $\rm H\rightarrow L$ counts versus
            probability $p$ for the probabilistic model. Top panel:
            Einstein-de~Sitter model; middle panel: open model;
            bottom panel: cosmological constant model.\label{fig5}}

\figcaption[]
{Time-evolution of the number density of galaxies around each
            elliptical galaxies described in the text as a 
            ``$\rm VL\rightarrow H$~elliptical,'' plotted as a function of
            redshift $z$, for all runs for which there
        are such galaxies. The number density $n$ is in units of ${\rm Mpc}^3$.
\label{fig6}}

\figcaption[]
{Location of the galaxies at $z=3$ for the EdS1 and EdS1$^*$ runs
            in computational units. (a)~$x$-coordinate of galaxies in the 
            EdS1$^*$ run versus same coordinates in the EdS1 run. 
            (b)~Histogram of the distance between each galaxy in the EdS1 
            run and its counterpart in the EdS1$^*$ run. 
            Each bin has a width of $800\,\rm kpc$ in physical units.
\label{fig7}}

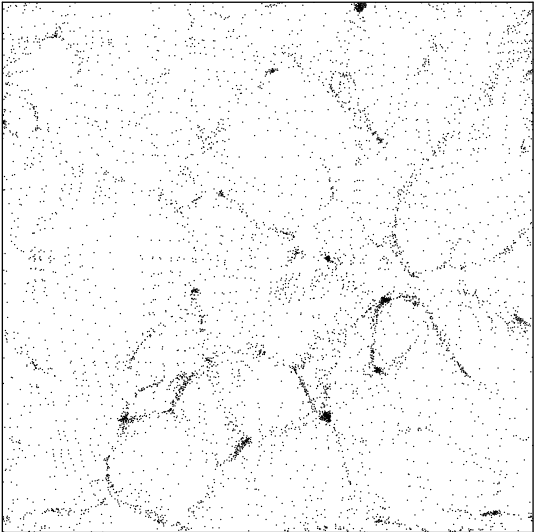
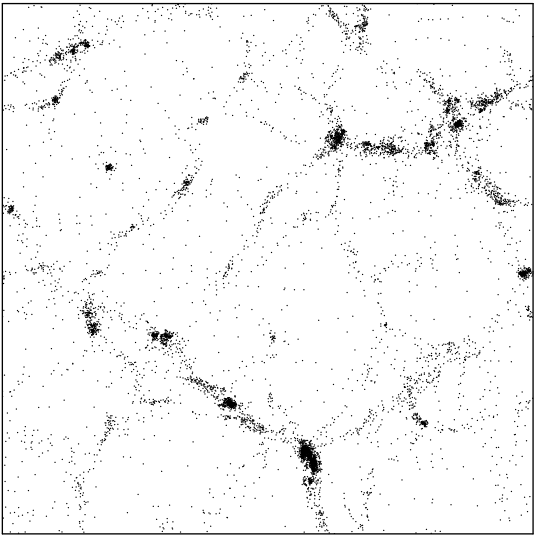
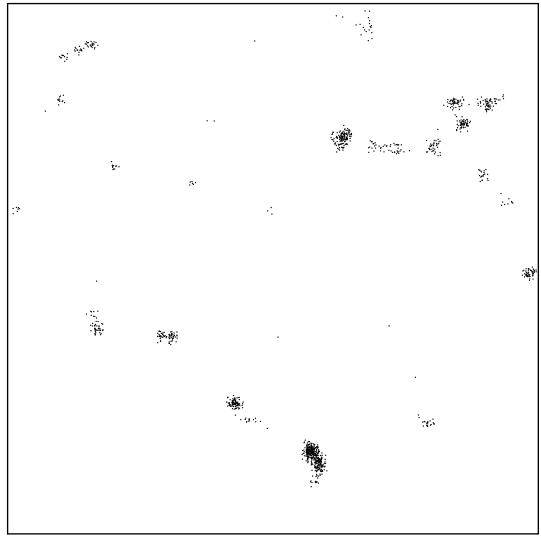
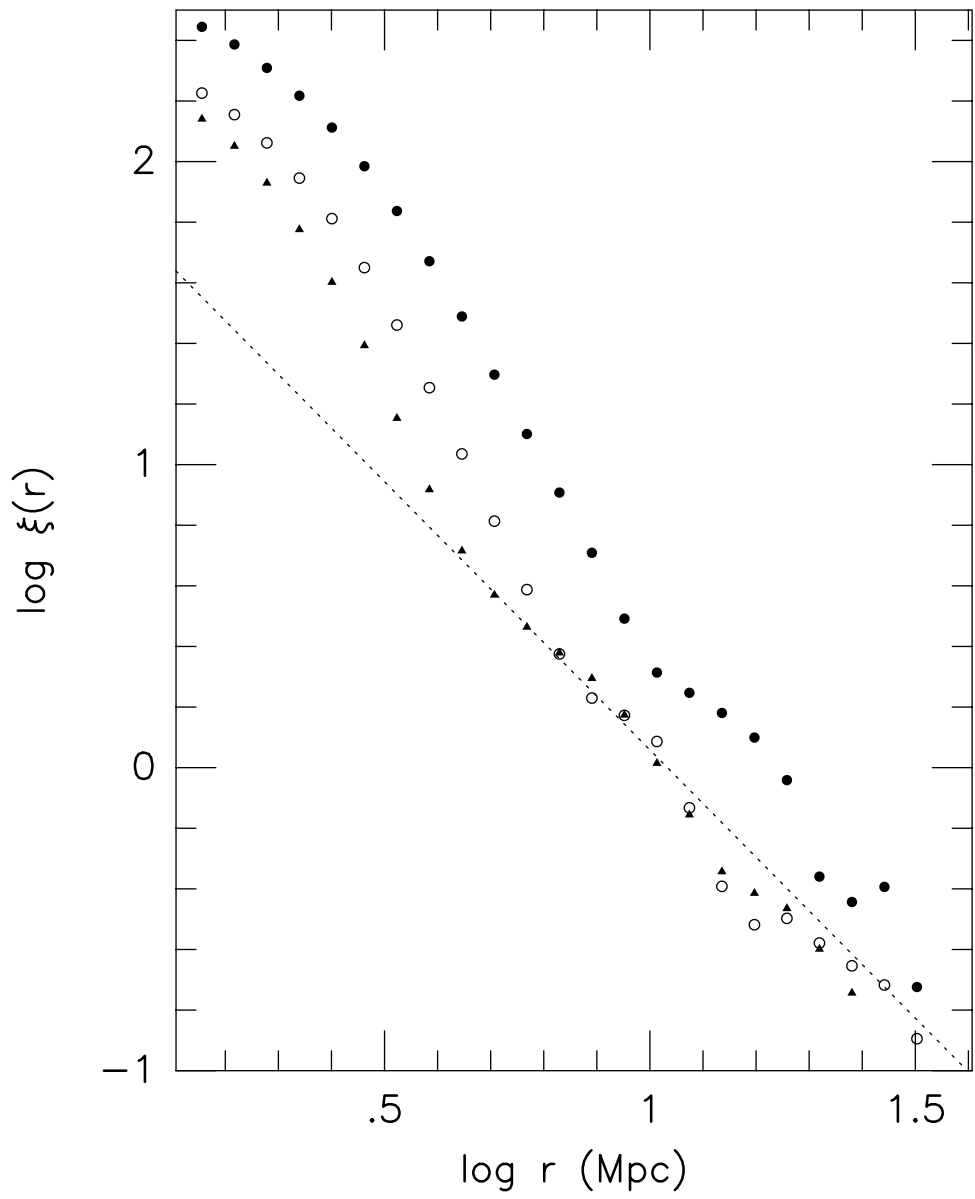
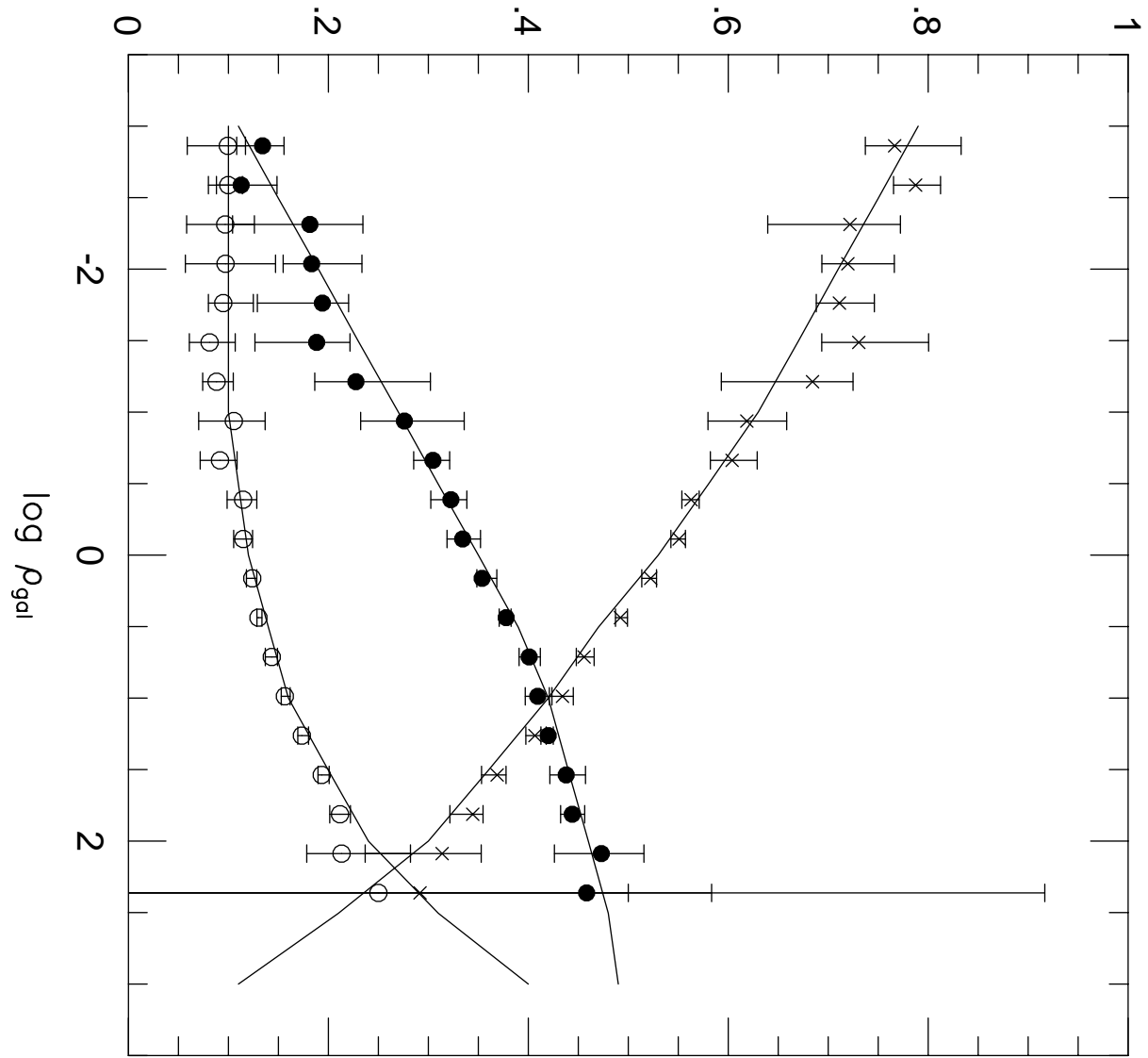
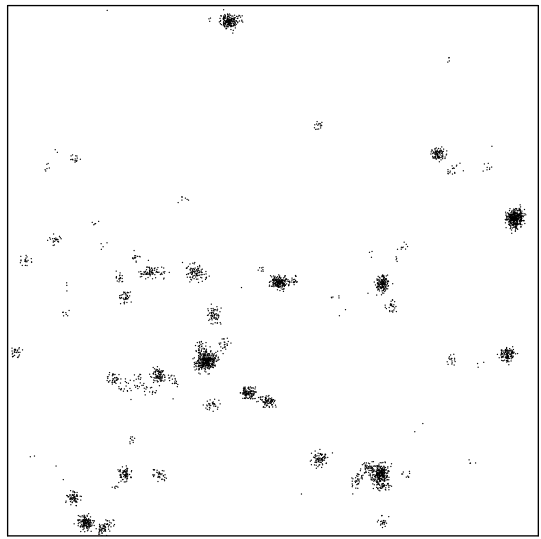
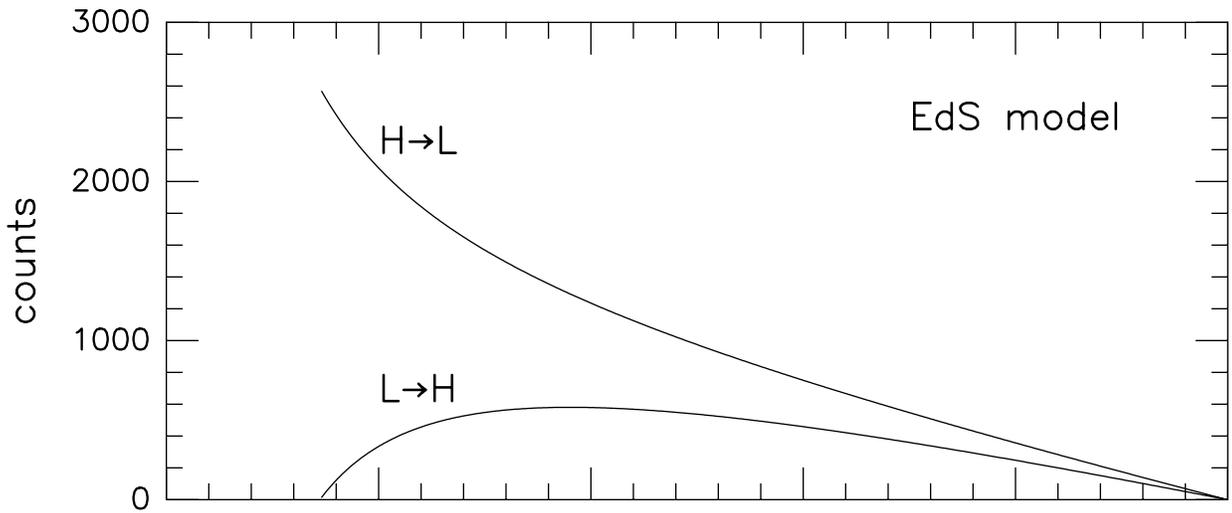
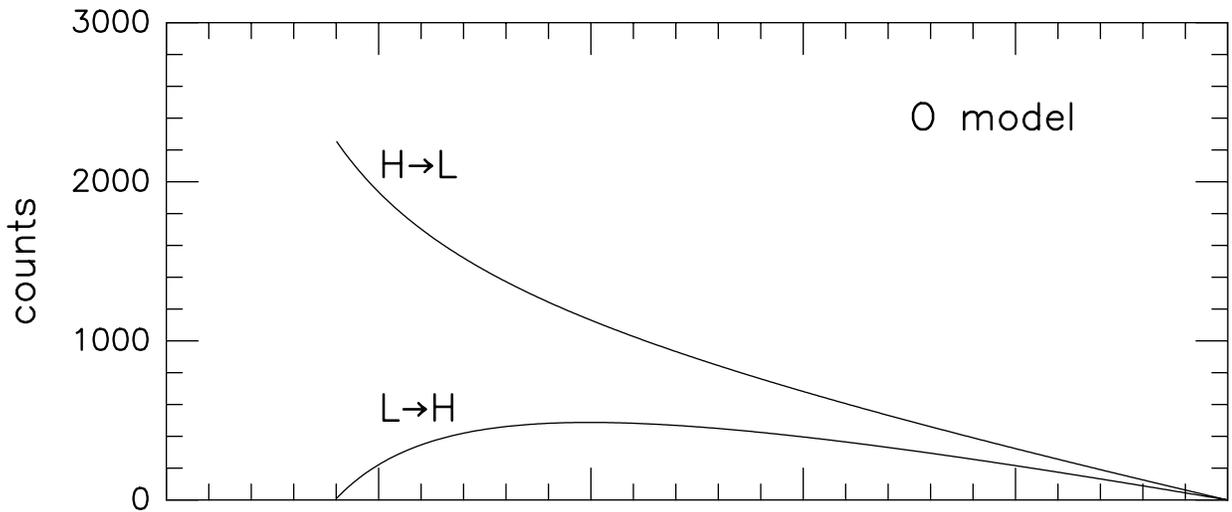
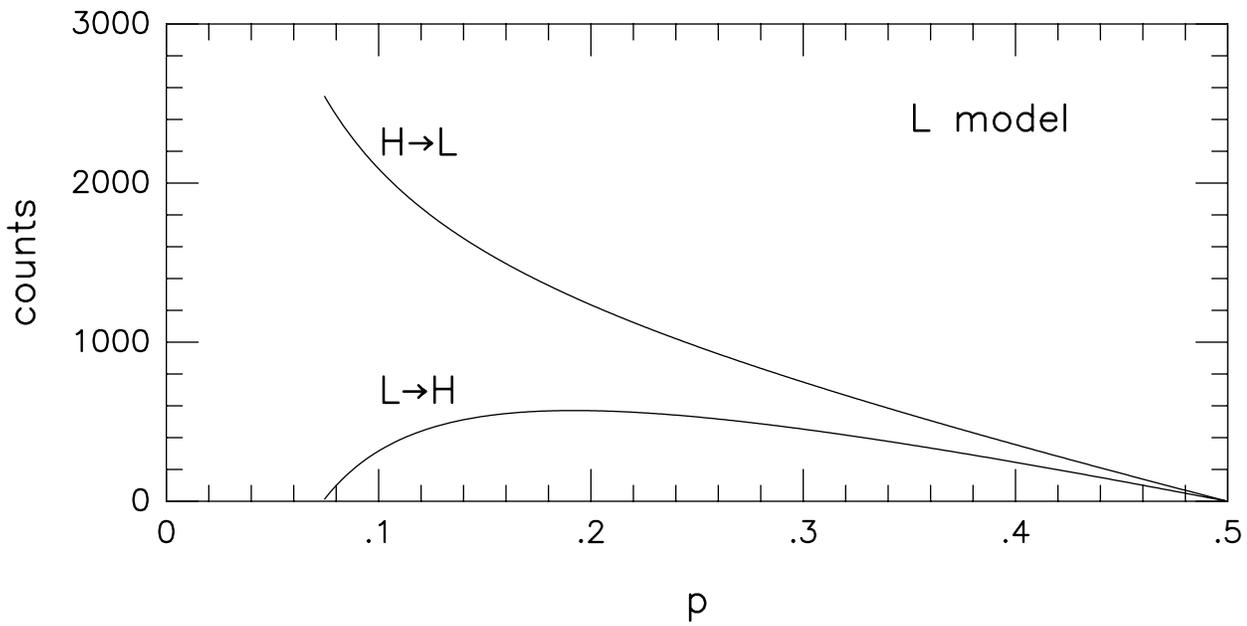
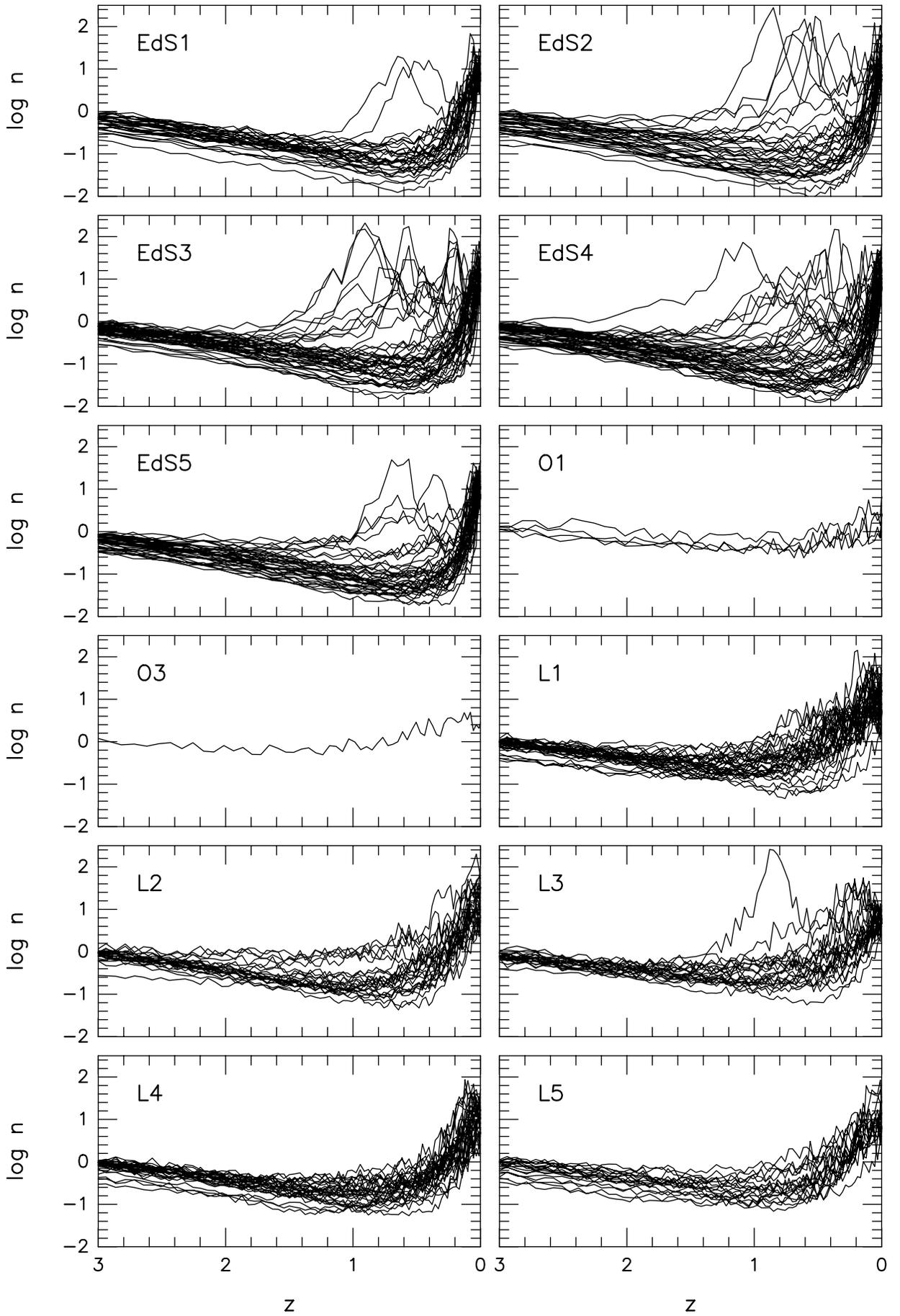
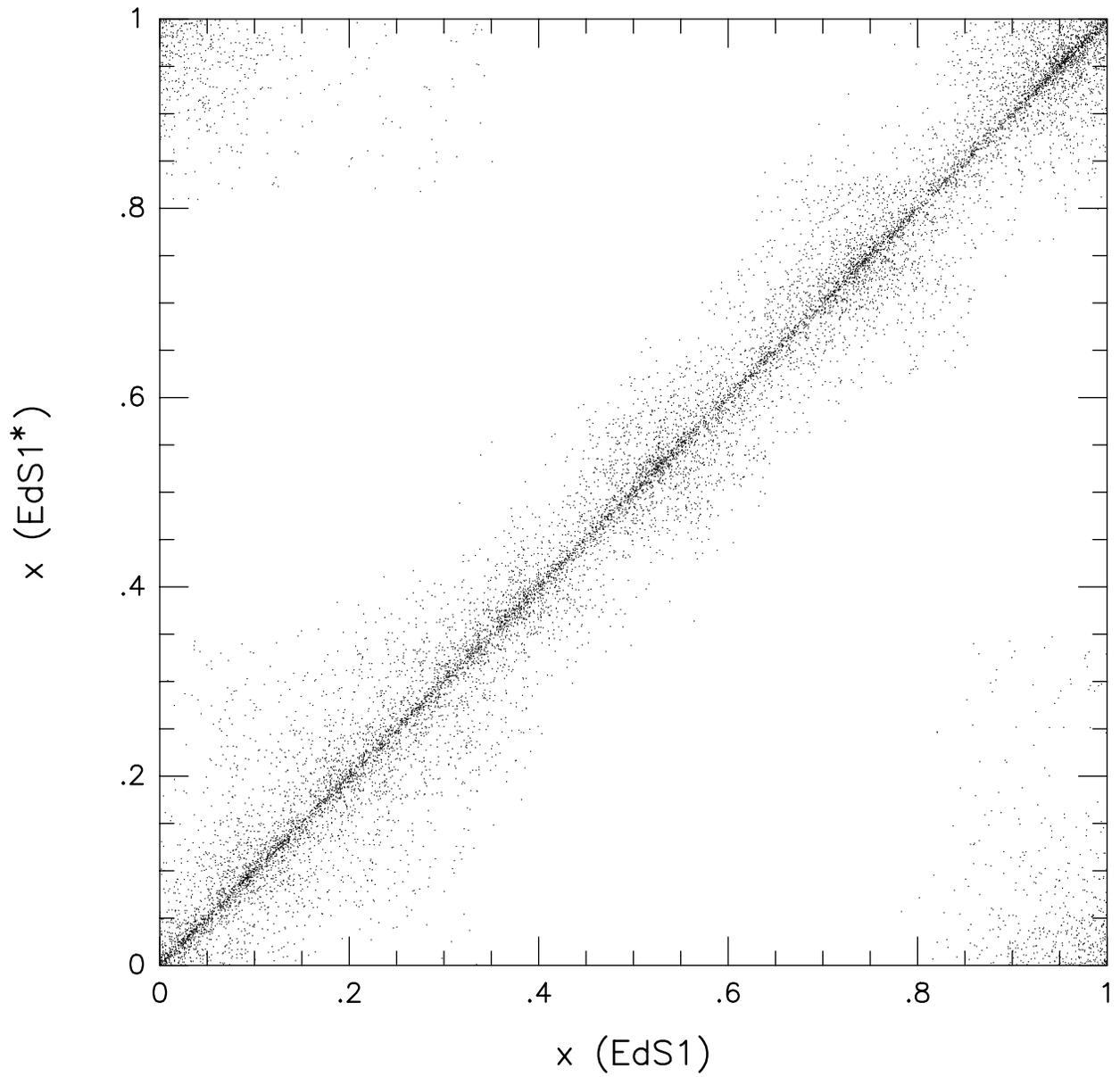
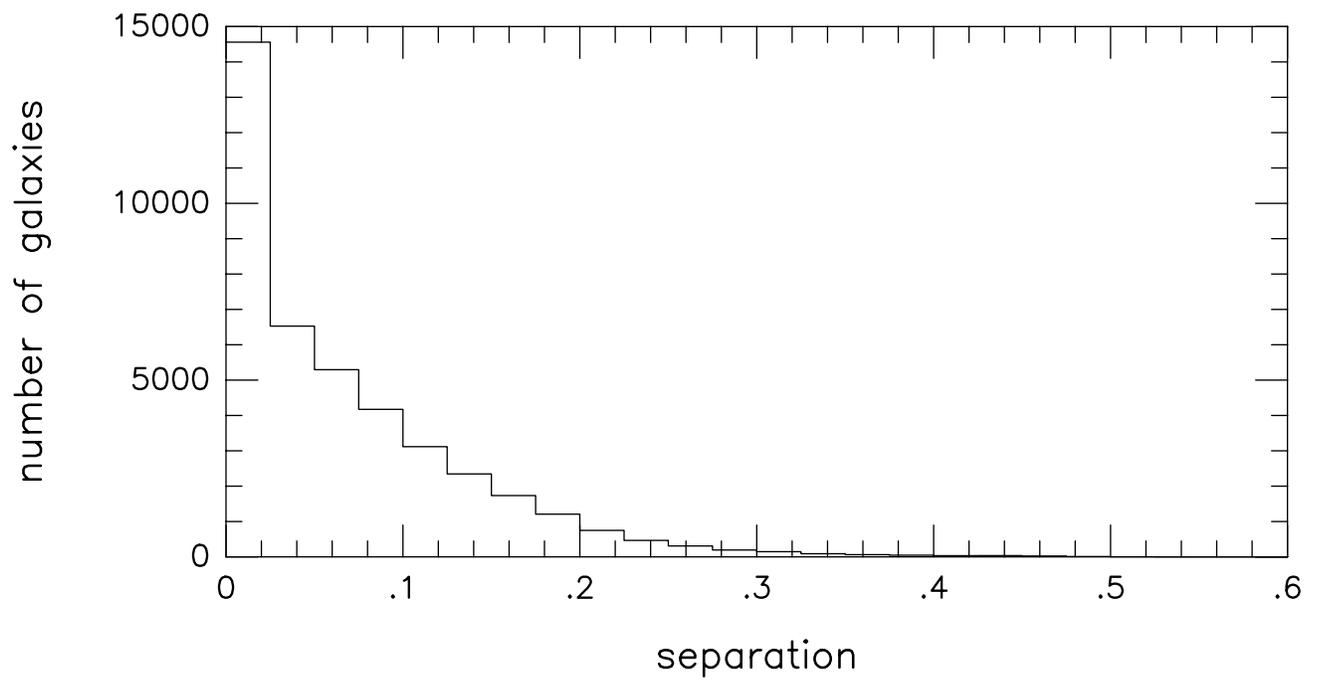
\end{document}